\documentclass[aip,
               preprint,
               amsmath,
               superscriptaddress,
               floatfix,
               nobibnotes,
              ]{revtex4-2}
              
\usepackage{amsfonts}
\usepackage{bm}
\usepackage[dvips]{graphicx}
\usepackage{color}
\usepackage[utf8]{inputenc}
\usepackage[flushleft]{caption}
\usepackage{subfigure}
\usepackage{rotating}
\usepackage{txfonts}
\usepackage{booktabs}
\usepackage[unicode]{hyperref}
\usepackage{xcolor}
\usepackage{indentfirst}
\usepackage{mathrsfs}  
\usepackage{braket}

\DeclareMathAlphabet{\mathpzc}{OT1}{pzc}{m}{it}

\def\figfoot{Phase Expert, Meng}
\def\mathbi#1{\textbf{\em #1}}
\def\mathfi#1{\boldsymbol{\mathfrak{#1}}}
\def\mi{\textrm{i}}

\newcommand{\figcaption}[2]{
    \noindent {\bf Figure \ref{#1}:} #2
    \vspace{1cm}
}

\begin{document}

\newtheorem{theorem}{Theorem}
\newtheorem{definition}{Definition}
\newtheorem{lemma}{Lemma}
\newtheorem{proposition}{Proposition}


\title{Berry Phase Effects of Nuclei in Chemical Reaction Dynamics}

\author{Xingyu Zhang}
 \affiliation{Department of Chemistry,
              Northwestern Polytechnical University,
              West Youyi Road 127, 710072 Xi'an,
              China}

\author{Jinke Yu}
 \affiliation{Department of Chemistry,
              Northwestern Polytechnical University,
              West Youyi Road 127, 710072 Xi'an,
              China}
              
\author{Qingyong Meng}
\email{qingyong.meng@nwpu.edu.cn}
 \affiliation{Department of Chemistry,
              Northwestern Polytechnical University,
              West Youyi Road 127, 710072 Xi'an,
              China}

\date{\today}


\begin{abstract}

\noindent {\bf Abstract}:
In calculations on quantum state-resolved dynamics of a chemical
reaction, reactants are usually prepared in separated eigenstates of
individual fragments, and their direct-product is then evolved in time.
In this work, we focus on the essence in separating them and the Berry
phase effects of the nuclear wave function. By the present theory,
mechanism of inter/intramolecular energy redistribution is also proposed
to deeply understand reactive dynamics with multirovibrational states.
To demonstrate the phase transition of the nuclear wave function, two
three-dimensional (3D) models reductively describing the molecular
reaction are developed to simulate transport of the system along a
closed path in a parameter space represented of inter/intramolecular
energy transfers. Employing these 3D models, extensive multiconfigurational
time-dependent Hartree (MCTDH) calculations are performed to solve the
time-dependent nuclear Schr{\"o}dinger equation at various initial
conditions. Moreover, 98D multilayer MCTDH (ML-MCTDH) calculations are
launched to demonstrate the transition of Berry phase. These
calculations clearly indicate that the wave function can change sign
allowing quantum interference in the parameter space. Discussions on
the separation of the reactants are made, while perspectives on the
Berry phase effects predicted by the present work are given from the
viewpoint of differential geometry. As a conclusion remark, the Berry
phase effects on molecular dynamics are also thoroughly compared with
those on electronic properties (see, for example, {\it Rev. Mod. Phys.}
{\bf 82} (2010), 1959) and mode/bond-specific reactivity (see, for
example, {\it Nat. Chem.} {\bf 14} (2022), 545).

\end{abstract}

\maketitle

\section{Introduction\label{sec:intro}}

Owing to the significant mass difference between electron and nuclear,
chemical dynamics typically begins by approximately separating electronic
and nuclear motions, called the Born-Oppenheimer approximation (BOA)
\cite{ced04:3,bae06:boa,ced13:224110}. It is the BOA which is not only
gives rise to adiabatic and diabatic representations of a molecular
system but also induces an additional phase in the electronic wave
function \cite{xia10:1959}, called Berry phase. The concept of Berry
phase is of immense interest across various fields of physics due to
its (1) gauge invariance, (2) geometrical essence, and (3) analogies
to gauge field theories and differential geometry. The Berry phase
effects \cite{ber84:45} lead subsequently to observable quantum
interference in chemical reaction. This is one of quantum effects that
are ubiquitous and crucial in understanding chemical reaction dynamics
\cite{yua18:1289,yan20:582,xie20:767,che21:936,wan21:938,pan22:545,wan23:191,li24:746,rei25:962}.
On the other hand, appropriate separation of degrees of freedom (DOFs)
and the subsequent dimensionality reduction play an important role in
quantum dynamics to significantly reduce computational costs. To do this,
the DOFs of a high-dimensional system are firstly divided into distinct
parts, such as fast or slow one, based on drastically different motion
rates or frequencies, and then either the slow or fast DOFs are fixed
at their optimal values according to chemical inspiration. For instance,
in quantum dynamics calculations on H + CH$_4$ $\to$ H$_2$ + CH$_3$,
the CH$_3$ fragment is usually fixed at the $C_{3v}$ symmetry. In
other words, the fast DOFs of CH$_3$, such as asymmetric C-H stretching
modes, are separated and frozen. Alternatively, in dynamics calculations
on surface scattering, the surface DOFs, {\it i.e.} the slow DOFs, are
often frozen. Such scheme of separation and reduction is also employed
in electronic structure where the nuclear coordinates are separated
due to the BOA.

In this work, we focus on the Berry phase of nuclear wave function
rather that that of electronic wave function. Over the past years,
separation of the molecule and the surface during the surface
scattering has been explored theoretically \cite{men15:164310,men22:16415},
implicitly producing the Berry phase of molecular wave function, where
the molecule and surface atoms are fast and slowly moving, respectively.
In this analysis, the adiabatic and diabatic representations of the
molecule-surface scattering were introduced. Here, we should mention
that the adiabatic representation of a process assumes that a system
undergoes a change without changing its surroundings, while the diabatic
representation is the opposite of the adiabatic one allowing the system
to exchange energy with its surroundings. Moreover, by the present
theory, a new mechanism is proposed in this wotk for inter/intramolecular
modes redistribution (IMR) which in a unique molecule also denotes
intramolecular vibrational energy redistribution (IVR). The IMR process
plays an important role in chemical dynamics, especially noting that
vibrational excitations often efficiently control chemical reactions
\cite{pol87:952,pol87:680}. A intuitive way based on dynamics calculations
could provide a definitive measure of this process by the time evolutions
of a limited number of accessible states. However, this way is often
too simplistic to fully understand IMR at full quantum-mechanical
fashion taking into account all the couplings. For instance, Pan and
Liu \cite{pan22:545} suggested that the scattering interferences
mediated by the CH$_3$D vibrational Fermi phases must be considered in
understanding its reaction with Cl. By the present theory, one can
construct a basis set of all possible rovibrational states (rather
than electronic states) to establish a non-adiabatic representation.
The dynamics calculations under this non-adiabatic representation will
explicitly reveal the IMR processes among different rovibrational states
during reaction.

The rest of this paper is organized as follows. Section \ref{sec:theo}
gives the present theory. Section \ref{sec:resu-diss} presents numerical
demonstrations, together with discussions and perspectives. Section
\ref{sec:con} concludes with a summary.

\section{Theory\label{sec:theo}}

\subsection{Adiabatic and Diabatic Representations\label{sec:adt}}

First of all, unless specified otherwise, the atomic units are always
employed in this work. In a formalist way, a two-body system can be
described by the point in the space $\mathbi{z}=\{\mathbi{q},\mathbi{Q},\boldsymbol{\lambda}\}$,
which is direct sum of a parameter space $\boldsymbol{\lambda}$ and the
configurational space $\{\mathbi{q},\mathbi{Q}\}$, where $\mathbi{q}$
and $\mathbi{Q}$ are coordinates of the two bodies. We further assume
that $\mathfi{z}=\mathbi{z}\setminus\mathbi{q}$. Letting the Hamiltonian
$\hat{H}(\mathbi{z})$ be explicitly time-independent, in adiabatic
evolution, where one of elements in $\mathbi{z}$ varies very slowly,
solution of the Schr{\"o}dinger equation $\mi\partial_t\Psi_n=\hat{H}\Psi_n$
is written as
\begin{equation}
\Psi_n\Big(\mathbi{z},t\Big)=\exp\Big(\mi\gamma_n(t)\Big)\exp\left[-
\mi\int_0^t\mathrm{d}t'\epsilon_n\Big(\mathbi{z}(t')\Big)\right]
\psi_n\Big(\mathbi{z}(0)\Big),
\label{eq:solution-of-se-000}
\end{equation}
where $n$ represents quantum numbers labeled the state. The function
$\psi_n(\mathbi{z}(0))$ is eigenstate of the Hamiltonian operator at
initial time $t=0$, while $\epsilon_n(\mathbi{z}(t))$ is the eigenvalue
corresponding to $\psi_n(\mathbi{z}(t))$, satisfying $\hat{H}\psi_n=\epsilon_n\psi_n$.
The second exponetial is known as the dynamical phase factor, while
$\gamma_n(t)$ is an additional phase. To understand $\gamma_n$,
substituting Equation \eqref{eq:solution-of-se-000} into the Schr{\"o}dinger
equation, the phase $\gamma_n$ is a path integral
\begin{equation}
\gamma_n=\mi\int_{\Gamma}\mathrm{d}\mathfi{z}\langle\psi_n\vert
\partial_{\mathfi{z}}\vert\psi_n\rangle_{\mathbi{q}}=
\int_{\Gamma}\mathrm{d}\mathfi{z}A_n(\mathfi{z}),
\label{eq:solution-of-se-001}
\end{equation}
where $\Gamma$ is integral path and
$A_n(\mathfi{z})=\mi\langle\psi_n\vert\partial_{\mathfi{z}}\vert\psi_n\rangle_{\mathbi{q}}$
is the Berry connection. For a general path, one can always choose an
appropriate gauge transform $\psi_n\to\psi_n\exp(\mi f)$ to cancel out
$\gamma_n$. 
For a closed path, because $\exp(\mi f)$ must be single valued, one
may not remove the phase
\begin{equation}
\gamma_n=\oint_{\Gamma}\mathrm{d}\mathfi{z}A_n\big(\mathfi{z}\big).
\label{eq:geome-phase-0000}
\end{equation}
In this case, $\gamma_n$ becomes gauge-invariant, known as the Berry
phase or geometric phase. If the above two-body system is a molecular
system, letting the nuclear coordiantes be slow DOFs, the above
derivations and definitions on the Berry phase and the Berry connection
lead to the Berry phase effects on electronic properties
\cite{ced04:3,bae06:boa,ced13:224110}. Furthermore, such derivations
have been used to demonstrate the existence of a phase in the molecular
wave function \cite{men22:16415} by optimizing factorized wave function
of the molecule-surface system \cite{ced13:224110,bae06:boa,men22:16415}.

To separate the DOFs of the above two-body system, let us derive
equations of motion (EOMs) for these bodies. Assuming that the sets of
mass-scaled coordinates $\mathbi{q}$ and $\mathbi{Q}$ describe the fast
and slow motions of a reaction system, respectively. The Hamiltonian
operator can be written as summation of the kinetic energy operator
(KEO) for the slow DOFs and the Hamiltonian operator for the fast
DOFs,
\begin{equation}
\hat{H}(\mathbi{z})=\hat{T}_{\mathrm{slow}}(\mathbi{Q},\boldsymbol{\lambda})
+\hat{H}_{\mathrm{fast}}'(\mathbi{z})=
\hat{T}_{\mathrm{slow}}(\mathbi{Q},\boldsymbol{\lambda})+
\hat{T}_{\mathrm{fast}}(\mathbi{q},\boldsymbol{\lambda})+
V_{\mathrm{fast}}(\mathbi{q},\mathbi{Q}_0,\boldsymbol{\lambda})+
V_{\mathrm{slow}}(\mathbi{Q},\boldsymbol{\lambda})
+V_{\mathrm{coul}}(\mathbi{z}),
\label{eq:wf-factorization-000}
\end{equation}
where $V_{\mathrm{fast}}$ is the potential energy surface (PES) for the
fast motions with the slow coordiantes fixed at $\mathbi{Q}_0$, while
$V_{\mathrm{slow}}$ is the PES for the slow motions. Moreover,
$V_{\mathrm{coul}}$ is the coupling term between the fast and slow
DOFs. Inspiring by factorization for the molecule-surface system
\cite{men22:16415}, the wave function of the entire system $\Psi(\mathbi{z})$
is factorized into a term $\varphi(\mathbi{z})$ for the fast motions
and a term $\chi(\mathfi{z})$ for the slow motions,
\begin{equation}
\Psi(\mathbi{z})=\varphi(\mathbi{z})\chi(\mathbi{Q},\boldsymbol{\lambda})
=\varphi(\mathbi{z})\chi(\mathfi{z}),
\label{eq:wf-factorization-002}
\end{equation}
where $\varphi$ depends parametrically on $\mathbi{Q}$ and is normalized
$\langle\varphi\vert\varphi\rangle_{\mathbi{q}}=1$, while we again
assume that $\mathfi{z}=\mathbi{z}\setminus\mathbi{q}$. Now, we should
optimize $\varphi$ and $\chi$ with normalization conditions
$\langle\varphi\vert\varphi\rangle_{\mathbi{q}}=1$ and
$\langle\Psi\vert\Psi\rangle_{\mathbi{q}}=1$ to improve the form of
the Hamiltonian operator of Equation \eqref{eq:wf-factorization-000}
and to derive the EOMs for $\varphi$ and $\chi$
such that $\Psi$ is automatically factorized as given in Equation
\eqref{eq:wf-factorization-002}. Letting $\mu_1$ and $\mu_2$ be
Lagrange parameters, the optimization target reads
\begin{equation}
\mathcal{J}\big[\varphi,\chi\big]=\big\langle\varphi\chi\big\vert
\hat{H}\big\vert\varphi\chi\big\rangle_{\mathbi{q}}+\mu_1\Big(1-
\big\langle\varphi\chi\big\vert\varphi\chi\big\rangle_{\mathbi{q}}\Big)+
\mu_2\Big(1-\big\langle\chi\big\vert\chi\big\rangle_{\mathbi{Q}}\Big).
\label{eq:wf-factorization-003}
\end{equation}
Varying $\mathcal{J}$ with respect to $\varphi$ and $\chi$ and setting
to zero, namely $\delta\mathcal{J}/\delta\varphi=0$ and $\delta\mathcal{J}/\delta\chi=0$,
one can obtain
\begin{equation}
\hat{H}_{\mathrm{fast}}\varphi-\big\langle\varphi\big\vert
\hat{H}_{\mathrm{fast}}\big\vert\varphi\big\rangle_{\mathbi{q}}\varphi
=\Big(\hat{H}_{\mathrm{fast}}-E\Big)\varphi=0,\quad
\Big(\hat{T}_{\mathrm{slow}}+E\Big)\chi=\mu_1\chi,
\label{eq:wf-factorization-011}
\end{equation}
where $E=\langle\varphi\vert\hat{H}_{\mathrm{fast}}\vert\varphi\rangle_{\mathbi{q}}$
is expectation of the effective Hamiltonian operator for the fast
motions,
\begin{equation}
\hat{H}_{\mathrm{fast}}=\hat{T}_{\mathrm{slow}}+
\hat{H}_{\mathrm{fast}}'-
\big(\boldsymbol{\nabla}_{\mathbi{Q}}\ln\chi\big)
\boldsymbol{\nabla}_{\mathbi{Q}}=\hat{H}-
\big(\boldsymbol{\nabla}_{\mathbi{Q}}\ln\chi\big)
\boldsymbol{\nabla}_{\mathbi{Q}}.
\label{eq:effective-ham-00}
\end{equation}
In deriving Equations \eqref{eq:wf-factorization-011} and \eqref{eq:effective-ham-00},
one can further find $\mu_2=0$. Thus, the separation of fast and slow
DOFs of the system is mathematically similar to that of the electrons
and nuclei in the molecular system. According to Equation \eqref{eq:effective-ham-00},
the effective Hamiltonian operator $\hat{H}_{\mathrm{fast}}$ has an
additional term $-(\boldsymbol{\nabla}_{\mathbi{Q}}\ln\chi)\boldsymbol{\nabla}_{\mathbi{Q}}$
that arises a phase in $\varphi$. To observe the phase term, we must
further consider the {\it adiabatic} and {\it diabatic} representations
of the entire system involved in multiple rovibrational states.

To this end, assuming $\varphi_j$ the $j$-th
eigenstate of $\hat{H}_{\mathrm{fast}}'$, one can extend Equation
\eqref{eq:wf-factorization-002} into the $N$-states expansion,
\begin{equation}
\Psi(\mathbi{z})=\sum_j^N\varphi_j(\mathbi{z})\chi_j(\mathfi{z}),
\label{eq:adi-dia-rep-000}
\end{equation}
where the $j$-th coefficient $\chi_j$ is the $j$-th element of the
representation of $\Psi$, that is matrix $\boldsymbol{\Psi}(\mathbi{z})$.
Substituting Equation \eqref{eq:adi-dia-rep-000} into the Schr{\"o}dinger
equation with $\hat{H}$, one can obtain
\begin{equation}
-\frac{1}{2}\nabla^2_{\mathfi{z}}\boldsymbol{\Psi}+\Big(\mathbi{E}-
\mu_1\mathbi{I}\Big)\boldsymbol{\Psi}-\frac{1}{2}\Big(
2\boldsymbol{\tau}\cdot\boldsymbol{\nabla}_{\mathbi{q}}+
\boldsymbol{\tau}^{(2)}\Big)\boldsymbol{\Psi}=\mathbf{0},
\label{eq:00-adi-dia-rep-001}
\end{equation}
where $\mathbi{E}$ and $\mu_1$ are matrix with elements $E_{ij}=
\langle\varphi_i\vert\hat{H}_{\mathrm{fast}}'\vert\varphi_j
\rangle_{\mathbi{q}}=E_i\delta_{ij}$ and eigenvalue of
$\hat{H}_{\mathrm{slow}}$, respectively, as shown by Equation
\eqref{eq:wf-factorization-011}. Moreover, the elements of non-adiabatic
coupling matrix (NACM) $\boldsymbol{\tau}$ and second-order NACM
$\boldsymbol{\tau}^{(2)}$ satisfy expressions
$\tau_{ij}=\langle\varphi_i\vert\nabla_{\mathfi{z}}\vert\varphi_j\rangle_{\mathbi{q}}$
and $\tau_{ij}^{(2)}=\langle\varphi_i\vert\nabla_{\mathfi{z}}^2\vert\varphi_j\rangle_{\mathbi{q}}$,
respectively. Noting that $\boldsymbol{\nabla}_{\mathbi{q}}\cdot\boldsymbol{\tau}=
\boldsymbol{\tau}^{(2)}-\boldsymbol{\tau}^2$, one can eliminate
$\boldsymbol{\tau}^{(2)}$ from Equation \eqref{eq:00-adi-dia-rep-001}
and obtain adiabatic EOM
\begin{equation}
-\frac{1}{2}\Big(\boldsymbol{\nabla}_{\mathfi{z}}+
\boldsymbol{\tau}\Big)^2\boldsymbol{\Psi}+\Big(\mathbi{E}
-\mu_1\mathbi{I}\Big)\boldsymbol{\Psi}=\mathbf{0}.
\label{eq:adi-dia-rep-001}
\end{equation}
Turning to the diabatic representation, the wave function is expanded
by
\begin{equation}
\tilde{\Psi}\big(\mathbi{z}\big)=\sum_j^N
\varphi_j\big(\mathbi{z}_0\big)\tilde{\chi}_j\big(\mathfi{z}\big),
\label{eq:adi-dia-rep-002}
\end{equation}
where $\mathbi{z}_0=\{\mathbi{q},\mathbi{Q}_0,\boldsymbol{\lambda}\}$
is obtained from $\mathbi{z}$ by fixing $\mathbi{Q}$ at $\mathbi{Q}_0$.
Similar to $\{\varphi_j(\mathbi{z})\}_{j=1}^N$ in expansion of Equation
\eqref{eq:adi-dia-rep-000}, the set $\{\varphi_j(\mathbi{z}_0)\}_{j=1}^N$
is an approximately complete set of eigenstates for $\hat{H}_{\mathrm{fast}}'(\mathbi{z}_0)$.
Substituting Equation \eqref{eq:adi-dia-rep-002} into the Schr{\"o}dinger
equation with $\hat{H}$, one can obtain diabatic EOM
\begin{equation}
-\frac{1}{2}\nabla_{\mathfi{z}}^2\tilde{\boldsymbol{\Psi}}+
\Big(\mathbi{V}-\mu_1\mathbi{I}\Big)\tilde{\boldsymbol{\Psi}}=
\mathbf{0}.
\label{eq:adi-dia-rep-006}
\end{equation}
where the diabatic potential matrix $\mathbi{V}(\mathfi{z})$ has the
element $V_{ij}(\mathfi{z})=\langle\varphi_i(\mathbi{z}_0)\vert
\hat{H}_{\mathrm{fast}}'(\mathbi{z})\vert\varphi_j(\mathbi{z}_0)
\rangle_{\mathbi{q}}$. Due to the difference between $\mathbi{z}_0$
and $\mathbi{z}$, the matrix $\mathbi{V}$ is non-diagonal.

\subsection{The Phase Effects of Nuclear Wavefunction\label{sec:phase-shift}}

The transform between the adiabatic and diabatic representations
requires transform matrix between $\boldsymbol{\varphi}(\mathbi{z})$
and $\boldsymbol{\varphi}(\mathbi{z}_0)$. Solving first-order
expansional expression
$\boldsymbol{\nabla}_{\mathbi{q}}\boldsymbol{\varphi}(\mathbi{z})+
\boldsymbol{\tau}\cdot\boldsymbol{\varphi}(\mathbi{z})=\mathbf{0}$
from $\mathbi{z}_1$ to $\mathbi{z}_2$, one can obtain
$\boldsymbol{\varphi}(\mathbi{z}_2)=\boldsymbol{\varphi}(\mathbi{z}_1)-
\int_{\mathbi{z}_1}^{\mathbi{z}_2}\mathrm{d}\mathbi{z}'\cdot
\boldsymbol{\tau}(\mathbi{z}')\cdot\boldsymbol{\varphi}(\mathbi{z}')$.
If $\mathbi{z}_1=\mathbi{z}_2$, the above integral along a closed path
$\Gamma$ in the space $\mathbi{z}$ is given by
\begin{equation}
\boldsymbol{\varphi}\big(\mathbi{z};\Gamma\big)=
\boldsymbol{\varphi}\big(\mathbi{z}\big)-
\oint_{\Gamma}\mathrm{d}\mathbi{z}'\cdot
\boldsymbol{\tau}(\mathbi{z}')\cdot
\boldsymbol{\varphi}(\mathbi{z}').
\label{eq:adi-dia-rep-010}
\end{equation}
Comparing with Equation \eqref{eq:geome-phase-0000}, the second term in
the right-hand side of Equation \eqref{eq:adi-dia-rep-010} can be
recognized as the Berry phase while the NACM $\mi\boldsymbol{\tau}(\mathbi{z})$
can be recognized as the Berry connection. To further explore this point,
Equation \eqref{eq:adi-dia-rep-010} is rewritten as
\begin{equation}
\boldsymbol{\varphi}\big(\mathbi{z};\Gamma\big)=
\mathscr{P}\exp\left(-\oint_{\Gamma}\mathrm{d}\mathbi{z}'\cdot
\boldsymbol{\tau}\big(\mathbi{z}'\big)\right)
\boldsymbol{\varphi}\big(\mathbi{z}),
\label{eq:adi-dia-rep-012}
\end{equation}
where $\mathscr{P}$ represents sequential integration carried out over
the space $\mathbi{z}$. Comparing Equation \eqref{eq:geome-phase-0000}
with Equation \eqref{eq:adi-dia-rep-012}, one can find
\begin{equation}
\varphi_j\big(\mathbi{z}_0;\Gamma)=\exp\Big(\mi\gamma_j\big(\Gamma\big)\Big)
\varphi_j\big(\mathbi{z}_0),\quad j=1,2,\cdots,N.
\label{eq:adi-dia-rep-014}
\end{equation}
Similar to the separation between electron and nuclear in the molecule,
if the parameter space $\mathbi{z}$ changes along the closed loop
$\Gamma$, the wave function for the fast motions undergoes a phase
change. For clarity, we compare the present phase effects with those
on electronic structure in Table \ref{tab:compare-phase}. The agreement
between both separation theories is evident that they both focus on the
dynamics of fast motions.

Equations \eqref{eq:adi-dia-rep-001} and \eqref{eq:adi-dia-rep-006}
give the time-independent EOMs of the adiabatic and diabatic
representations, respectively. If both $\boldsymbol{\Psi}$ and
$\tilde{\boldsymbol{\Psi}}$ are explicitly time-dependent, one
can similarly and directly extend the time-independent EOMs to,
\begin{equation}
\mi\frac{\partial}{\partial t}\boldsymbol{\Psi}(t)=\left[
-\frac{1}{2}\Big(\boldsymbol{\nabla}_{\mathfi{z}}
+\boldsymbol{\tau}\Big)^2+\mathbi{E}\right]\boldsymbol{\Psi}(t)=
\mathbi{H}_{\mathrm{adia}}\boldsymbol{\Psi}(t),\quad
\mi\frac{\partial}{\partial t}\tilde{\boldsymbol{\Psi}}(t)=
\left(-\frac{1}{2}\nabla_{\mathfi{z}}^2+\mathbi{V}\right)
\tilde{\boldsymbol{\Psi}}(t)=\mathbi{H}_{\mathrm{dia}}
\tilde{\boldsymbol{\Psi}}(t),
\label{eq:new-adi-dia-rep-016}
\end{equation}
where $\mathbi{H}_{\mathrm{adia}}$ and $\mathbi{H}_{\mathrm{dia}}$ are
time-independent Hamiltonian matrices of the adiabatic and diabatic
representations, respectively. By solving Equation
\eqref{eq:new-adi-dia-rep-016}, one can elucidate the mechanism of
IMR, say the IVR process, which is analogous to understand how a
molecule evolves among various electronic states. We must mention
that the relation between $\mathbi{E}$ and $\mathbi{V}$ is usually
unknown. To overcome this problem, we approximately rewrite Equation
\eqref{eq:new-adi-dia-rep-016} as
\begin{equation}
\mi\frac{\partial}{\partial t}\tilde{\boldsymbol{\Psi}}(t)=
\left(-\frac{1}{2}\nabla^2_{\mathfi{z}}+\mathbi{U}\right)
\tilde{\boldsymbol{\Psi}}(t)=\mathbi{H}_{\mathrm{nonadia}}
\tilde{\boldsymbol{\Psi}}(t).
\label{eq:new-adi-dia-rep-017}
\end{equation}
where $\mathbi{H}_{\mathrm{nonadia}}$ and $\mathbi{U}$ mean nonadiabatic
Hamiltonian and potential matrices, respectively, of the fast motions.
To construct $\mathbi{U}$, appropriate models must be developed to
account for inter-state couplings during IMR. Similar to Equation
\eqref{eq:solution-of-se-000}, since $\mathbi{H}_{\mathrm{adia}}$ and
$\mathbi{H}_{\mathrm{dia}}$ are both time-independent, propating the
$j$-th state of the system along the closed path $\Gamma$ in the parameter
space, one can obtain
\begin{equation}
\tilde{\Psi}_j\big(\mathbi{z}_0,t;\Gamma\big)=
\tilde{\Psi}_j\big(\mathbi{z}_0,t\big)
\exp\big(i\gamma_j(t;\Gamma)\big),
\label{eq:td-wp-adi-dia-001}
\end{equation}
where the dynamical phase factor has been absorbed into the
$\tilde{\Psi}_j(\mathbi{z}_0,t)$ term. According to Equation
\eqref{eq:td-wp-adi-dia-001}, the autocorrelation function for the fast
motions undergoes a phase.

Next, let us turn to the effect of the parameter set $\boldsymbol{\lambda}$
and introduce the concept of so-called pathological point in the
configurational space. The core idear is that the inequality
$\oint_{\Gamma}\mathrm{d}\mathbi{z}'\cdot\boldsymbol{\tau}(\mathbi{z}')\neq0$
does not imply the existence of
$\oint_{\Gamma}\mathrm{d}\mathbi{Q}'\cdot\boldsymbol{\tau}(\mathbi{z}')\neq0$.
Thus, only integral along the closed path in the configurational space
$\mathbi{Q}$ may not indicate a nonzero phase. Moreover, it is possible
to find a pathological point in the configurational space such that
$\oint_{\Gamma}\mathrm{d}\mathbi{Q}'\cdot\boldsymbol{\tau}(\mathbi{z}')\neq0$
when $\Gamma$ encircles that pathological point. As given by Table
\ref{tab:compare-phase}, the pathological point in the eletronic-structure
theory is conical intersection (CI) between adjacent electronic states.
To topologically illustrate the CI point, the PESs should be expressed
as a function of the branch space $g$-$h$ near CI, where the gradient
difference $g$ describes the gradient difference between the two
electronic states, while the non-adiabatic coupling $h$ describes the
strength of coupling. Both $g$ and $h$ can be seen as parameters of
electronic structure.
For the present theory on separation of the fast and slow DOFs in the
reaction, a similar pathological point in the configurational space
might not exist. However, it is still possible to find
$\oint_{\Gamma}\mathrm{d}\mathbi{z}'\cdot\boldsymbol{\tau}(\mathbi{z}')\neq0$
where $\Gamma$ can be find in the parameter space $\boldsymbol{\lambda}$
that represents environmental conditions of the reaction. For instance,
we previously proposed two-dimensional coupled harmonic oscillators
(2D CHO) model \cite{men15:164310,men17:184305} to explain the coupling
in the molecule-surface scattering, where the masses and frequencies
are $\{m,M\}$ and $\{\omega,\Omega\}$, respectively, while the ratio
$\Omega/\omega$ represents the surface temperature. As previously
discussed \cite{men15:164310,men17:184305}, the 2D CHO model predicts
that at $\Omega/\omega=\sqrt{1-m/M}$ an assignment between two oscillators
becomes unclear, rather than individual motion, called critical point.
This is similar to the
situation where the reaction occurs very fastly through the CI point
and thus motions of the electrons and nuclei can not be separately
considered at CI. It is also clear that the IER (or IVR) process
requires the aid of the slow motions. It should be mentioned that the
IER process is arisen from the couplings between the modes associated
with the slow motions, such as the surface motions in surface scattering.
For example, as expected by quantum dynamics calculations, the surface
phonons play an central role in the IVR process of the molecule-surface
scattering.

\section{Results and Discussions\label{sec:resu-diss}}

\subsection{Models for Chemical Reaction\label{sec:num-details}}

\begin{figure}[h!]
   \centering
     \includegraphics[width=18cm,angle=-90.0]{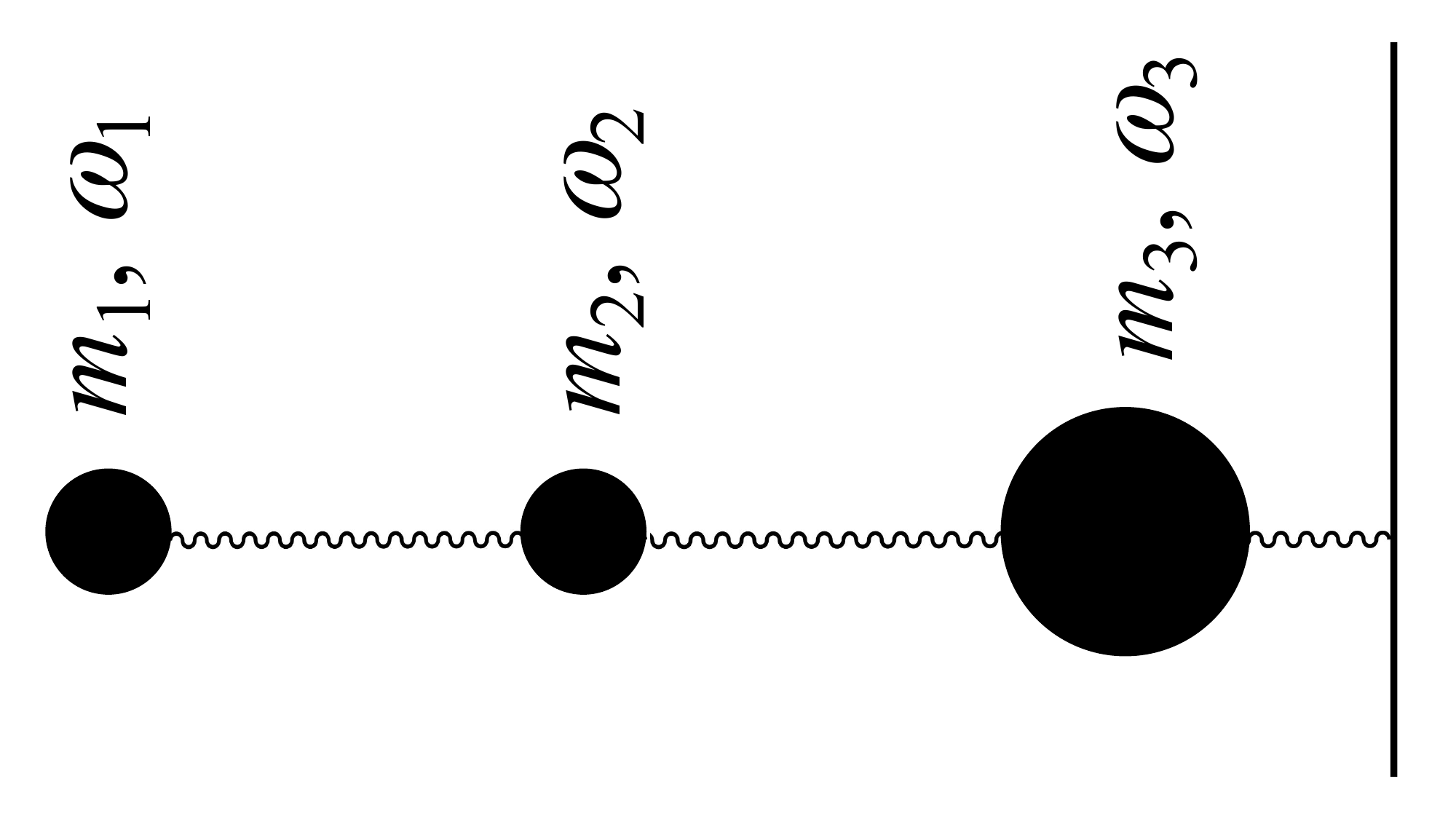}
     \caption{\figfoot}
      \label{fig:cho}
       \end{figure}

To demonstrate the present theory for chemical reactions, we firstly
introduce two numerical models for the simplest three-body reactions
and surface scattering. By these models, numerical demonstrations on
the present theory will be given and discussed. The first one is model
for surface scattering of a diatomic molecule. As previously discussed
\cite{men15:164310,men22:16415} on the 2D CHO model for the absorbed
molecule on the surface, the avoided crossing scenario of the vibrational
eigenenergies in frequencies space was demonstrated, where the surface
atom is bound harmonically to its equilibrium position and the molecule
is bound harmonically to the surface atom. However, the 2D CHO model
\cite{men15:164310,men22:16415} cannot simulate internal motions of
the adsorbed molecule. To overcome this problem, the 2D CHO model
\cite{men15:164310,men22:16415} is extended to a three-dimensional
(3D) CHO model, as illustrated in Figure \ref{fig:cho}, where
$\{m_i\}_{i=1}^3$ and $\{q_i\}_{i=1}^3$ represent mass and cooridnate
sets of the oscillators, respectively, while $\{\omega_i\}_{i=1}^3$
denotes associated frequencies set. Figure \ref{fig:cho} implies that
$\omega_1$ and $\omega_3$ model the intramolecular potentials, while
$\omega_2$ models the molecule-surface potential. According to Figure
\ref{fig:cho}, the 3D CHO Hamiltonian reads
\begin{align}
2\hat{H}_{\mathrm{3D-CHO}}\big(q_1,q_2,q_3,\eta,\xi\big)&=
\frac{\hat{p}_1^2}{m_1}+\frac{\hat{p}_2^2}{m_2}+\frac{\hat{p}_3^2}{m_3}+
m_3\omega_3^2q_3^2+m_1\omega_1^2(q_1-q_2)^2+m_2\omega_2^2(q_2-q_3)^2
\allowdisplaybreaks[4] \nonumber  \\
&=\hat{\tilde{p}}_1^2+\hat{\tilde{p}}_2^2+\hat{\tilde{p}}_3^2+
\omega_3^2\tilde{q}_3^2+\omega_1^2\left(\tilde{q}_1-\zeta_1\tilde{q}_2\right)^2
+\omega_2^2\left(\tilde{q}_2-\zeta_2\tilde{q}_3\right)^2
\allowdisplaybreaks[4] \nonumber  \\
&=\hat{\tilde{p}}_1^2+\hat{\tilde{p}}_2^2+\hat{\tilde{p}}_3^2+\omega_2^2
\left(\begin{array}{ccc}
\tilde{q}_1 & \tilde{q}_2 & \tilde{q}_3
\end{array} \right)
\left(\begin{array}{ccc}
\eta         & -\zeta_1\eta    & 0  \\
-\zeta_1\eta & 1+\zeta_1^2\eta & -\zeta_2  \\
0            & -\zeta_2        & \xi+\zeta_2^2
\end{array}\right)
\left(\begin{array}{c}
\tilde{q}_1  \\
\tilde{q}_2  \\
\tilde{q}_3
\end{array}\right),
\label{eq:3d-cho-hamiltonian-00}
\end{align}
where $\tilde{q}_i=\sqrt{m_i}q_i$ and $\tilde{p}_i=p_i/\sqrt{m_i}$ with
$i=1,2,3$, are mass-scaled coordinates and momenta, respectively. We
also define the frequency parameters as $\eta=\omega_1^2/\omega_2^2$
and $\xi=\omega_3^2/\omega_2^2$ and coupling paramters as $\zeta_1=
\sqrt{m_1/m_2}$ and $\zeta_2=\sqrt{m_2/m_3}$. The set of $\{\xi,\eta\}$
is thus the parameter space of the 3D CHO model. The frequencies of the
normal mode are related to the eigenvalues of the $3\times3$ potential
matrix in Equation \eqref{eq:3d-cho-hamiltonian-00}. Unlike the 2D CHO
model that exhibits an avoided crossing region \cite{men15:164310,men22:16415},
eigenvalues of the 3D CHO model as functions of $\eta$ and $\xi$ have
degeneracy seams, whose projection on $\eta$ is critical point $\eta_c$.
Analyzing degeneracy of the eigenvalues, the critical point $\eta_c$
only depends on the mass of the fragments, namely
$\eta_c=\sqrt{1+m_1/\max(m_2,m_3)}$. This implies that the degeneracy
feature is an intrinsic properties of the system. Such discrepancy
between avoided crossing and degeneracy seam is arisen from that two
restrictions on the system must be simultaneously satisfied for
existence of degenerated states. Since eigenstate of the 2D CHO model
\cite{men15:164310,men22:16415} has only one variable, that is frequency
ratio of two oscillators, its eigenstates cannot be degenerate. The
situation of the 3D CHO model changes due to its two variables, $\eta$
and $\xi$, where its eigenstates are possibly degenerated at one-dimensional
(1D) seam lines or zero-dimensional seam points. We would like to finally
emphasize that Equation \eqref{eq:3d-cho-hamiltonian-00} holds by expanding
the three-body interactions to second-order Taylor expansions with omission
of anharmonic contributions and assumptions of negligible interaction
between $m_1$ and $m_3$.

\begin{figure}[h!]
   \centering
    \subfigure[\quad Surface plot and contour plots of the PES for the 3D REX model at $\theta=\pi/2$]{
     \includegraphics[width=18cm]{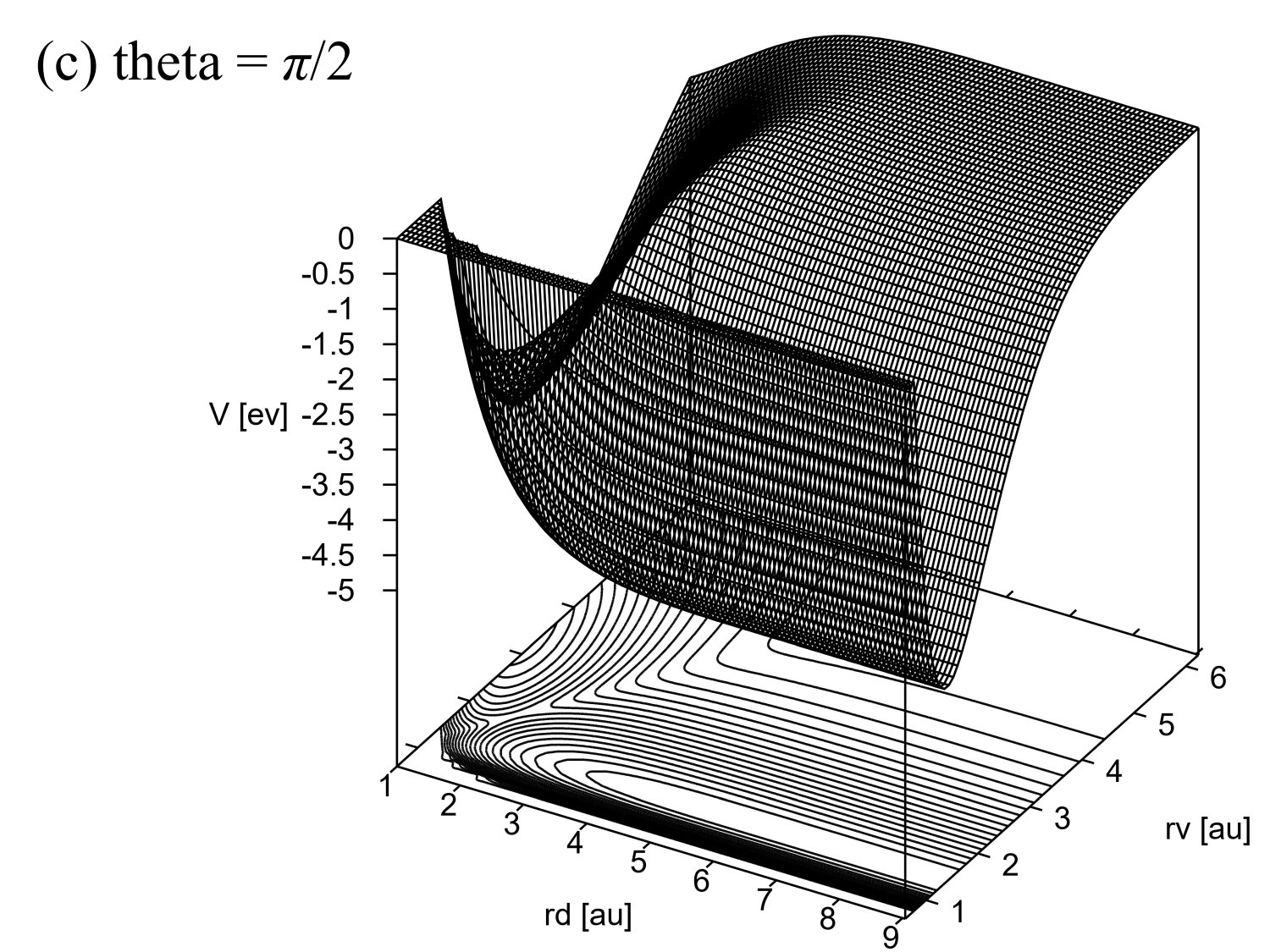}}
       \caption{\figfoot}
        \label{fig:rex-pes}
         \end{figure}

Next, we turn to model for a bimolecular reaction (REX) which can be
reduced to a simple three-fragment exchange process described by the
Jacobi cooridantes set $\{r_{\mathrm{v}},r_{\mathrm{d}},\theta\}$,
called 3D REX model. Three fragments labeled by $\#1$, $\#2$, and $\#3$
have masses $\{m_i\}_{i=1}^3$, where the fragment $\#1$ is
the colliding group and the fragments $\#2$ and $\#3$ the target. As
the reaction coordinates, vector $\vec{r}_{\mathrm{v}}$ is from $m_2$
to $m_3$, $\vec{r}_{\mathrm{d}}$ from $m_1$ to the center-of-mass (COM)
of the other two fragments, while $\theta$ is the angle between
$\vec{r}_{\mathrm{v}}$ and $\vec{r}_{\mathrm{d}}$. For simple, the
total rotations of the entire system is ignored by setting the total
angular momentum to be zero. With the above assumptions, the PES of
the 3D REX model reads $V_{\mathrm{3D-REX}}=V_3+V_{\mathrm{rev}}$. The
$V_3$ term is a 3D model potential along the reaction cooridnates based
on the BKMP2 PES of the H + H$_2$ system \cite{boo96:7139} and produces
main features fo the bimolecular reaction. The $V_{\mathrm{rev}}$ term
is the revision and perturbation due to the coupling between the reaction
coordinate and the others. Similar to the derivation of Equation
\eqref{eq:3d-cho-hamiltonian-00}, assuming that interactions between
each pairs of fregments are truncated at the second order, we have
\begin{equation}
2V_{\mathrm{rev}}\big(r_{\mathrm{v}},r_{\mathrm{d}},\theta,\alpha,\beta\big)=
\omega_{23}^2\left[\Big(\mu_{23}+\kappa_1\alpha+\kappa_2\beta\Big)
r_{\mathrm{v}}^2+\Big(\mu_{12}\alpha+\mu_{13}\beta\Big)r_{\mathrm{d}}^2
-\frac{1}{\nu_{23}}\Big(\kappa_1m_2\alpha-\kappa_2m_3\beta\Big)
r_{\mathrm{v}}r_{\mathrm{d}}\cos\theta\right],
\label{eq:reax-3d-rex-0000}
\end{equation}
where
\begin{equation}
\mu_{12}=\frac{m_1m_2}{m_1+m_2},\quad\mu_{13}=\frac{m_1m_3}{m_1+m_3},\quad
\mu_{23}=\frac{m_2m_3}{m_2+m_3},\quad
\kappa_1=\mu_{12}\left(\frac{\mu_{23}}{m_2}\right)^2,\quad
\kappa_2=\mu_{13}\left(\frac{\mu_{23}}{m_3}\right)^2.
\label{eq:reax-3d-rex-0001}
\end{equation}
Letting $\{\omega_{ij}\}_{i,j=1}^3$ be frequencies between two distinct
fragments, the parameters $\alpha$ and $\beta$ in Equation \eqref{eq:reax-3d-rex-0000}
satisfy
\begin{equation}
\alpha=\frac{\omega_{12}^2}{\omega_{23}^2},\quad
\beta=\frac{\omega_{13}^2}{\omega_{23}^2}.
\label{eq:reax-3d-rex-0002}
\end{equation}
Therefore, the Hamiltonian of the present 3D REX model is given by
\begin{equation}
2\hat{H}_{\mathrm{3D-REX}}\big(r_{\mathrm{v}},r_{\mathrm{d}},\theta,\alpha,\beta\big)
=-\frac{1}{\mu}\frac{\partial^2}{\partial r_{\mathrm{d}}^2}-
\frac{1}{\mu_{23}}\frac{\partial^2}{\partial r_{\mathrm{v}}^2}-
\left(\frac{1}{\mu r_{\mathrm{d}}^2}+\frac{1}{\mu_{23}r_{\mathrm{v}}^2}\right)
\frac{1}{\sin\theta}\frac{\partial}{\partial\theta}\sin\theta\frac{\partial}{\partial\theta}
+2V_{\mathrm{3D-REX}}\big(r_{\mathrm{v}},r_{\mathrm{d}},\theta,\alpha,\beta\big),
\label{eq:reax-3d-rex-0003}
\end{equation}
where $\mu=m_1(m_2+m_3)/(m_1+m_2+m_3)$ is the reduced mass associated
with $r_{\mathrm{d}}$. As illustrated
by the surface plots and contour plots of $V_{\mathrm{3D-REX}}$ in Figure
\ref{fig:rex-pes}, the term $V_{\mathrm{rev}}$ in Equation \eqref{eq:reax-3d-rex-0000}
is set to be a perturbation of $V_3$ and thus $V_{\mathrm{3D-REX}}=V_3+V_{\mathrm{rev}}$
is approximately a mass model of chemical reaction. 

\begin{figure}[h!]
 \centering
  \subfigure[\quad Phase angle of the 3D REX model]{
   \includegraphics[width=18cm]{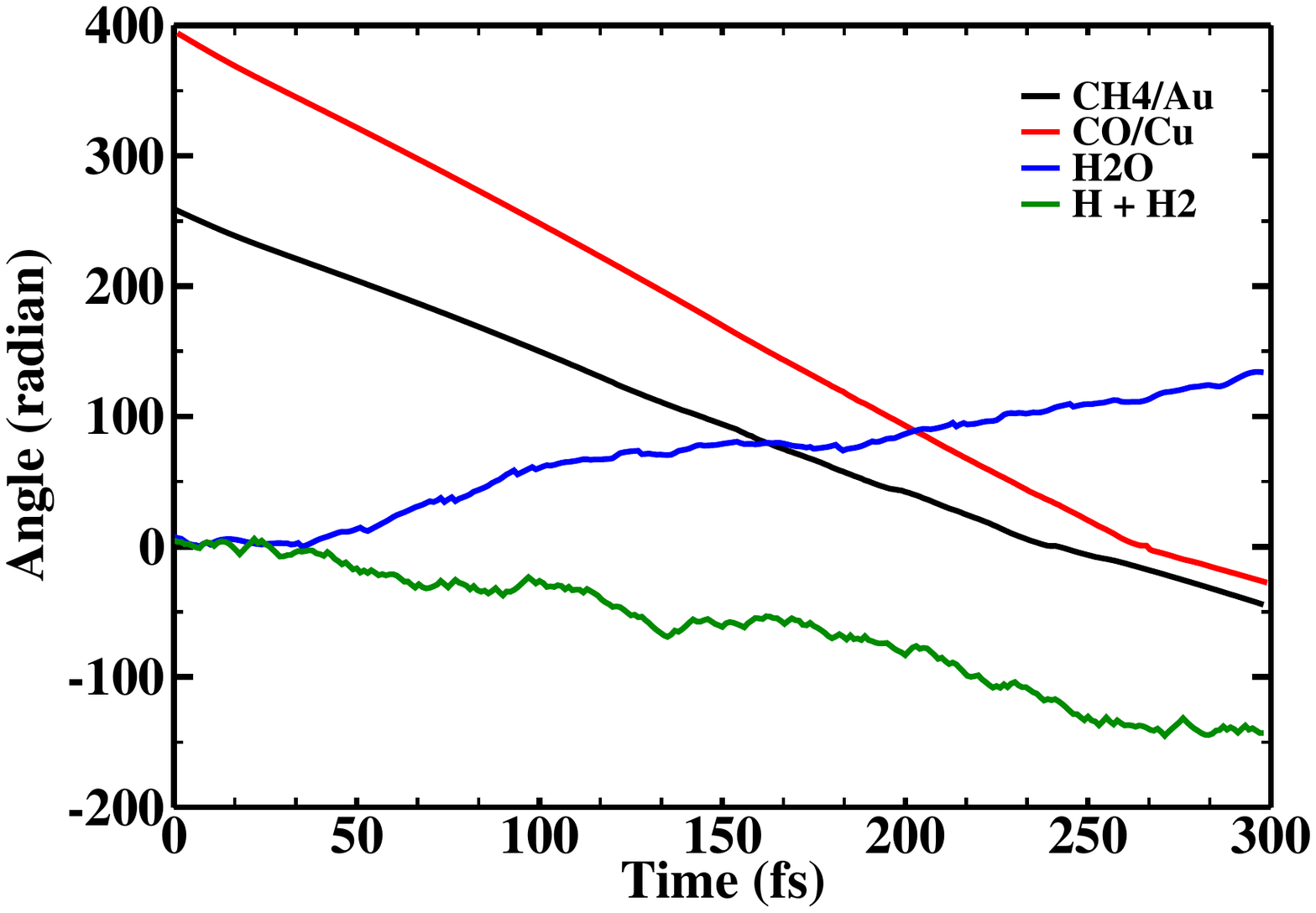}
}  
     \caption{\figfoot}
      \label{fig:tot-phase}
       \end{figure}

Collected in Tables \ref{tab:mass-model} and \ref{tab:model} are
parameters of the present model Hamiltonian operators. Given in Table
\ref{tab:num-prop} are primitive basis functions of discrete variable
representation (DVR) \cite{bac89:469,suk01:2604}, together with the
number of the grid points and the range of the grids which are all
used for representing the present calculations through the
multiconfigurational time-dependent Hartree (MCTDH) method
\cite{wan03:1289,man08:164116,ven11:044135,wan15:7951,mctdh:MLpackage}.
According to Equations \eqref{eq:3d-cho-hamiltonian-00} and
\eqref{eq:reax-3d-rex-0003}, almost all of KEO and PES terms
are already given in a sum-of-products (SOP) form except for the
$V_3$ term. Since the MCTDH calculations need the Hamiltonian operator
in the SOP form, the potential re-fitting (POTFIT) calculations
\cite{jae96:7974,jae98:3772} have to be firstly performed to transfer
the $V_3$ term into the SOP form. We refer the reader to Reference
\cite{son19:114302} for numerical details of the POTFIT calculations
on the H + H$_2$ system. The re-fitting error of present POTFIT
calculation is smaller than $1\:\mathrm{meV}$ ensuring accuracy of
subsequent MCTDH calculations. We refer the reader to References
\cite{mey90:73,bec00:1,mey09:book,mey12:351,mey18:124105,wan18:044119}
for further details on the MCTDH and POTFIT methods. By propagated wave
function $\Psi(t)$, one can compute time-dependent phase angle for
various models, as illustrated by Figure \ref{fig:tot-phase}. According
to Equation \eqref{eq:solution-of-se-000}, the phase of $\Psi(t)$ should
be linear function of time if energy is conserved and there is no
transition of the phase angle. Therefore, a slight deviation from the
linear relationship in Figure \ref{fig:tot-phase}, in particular that
in Figure \ref{fig:tot-phase}(b) for 3D REX model, implies changes in
energy during the reaction process as well as phase changes during the
parameter variation. We shall return to this point later after show the
transition of phase angle at the critical point.

\subsection{Numerical Demonstrations\label{sec:results-num}}

Figure \ref{fig:phase} illustrates time-dependent phase angle of the
nuclear wave functions computed by MCTDH with the 3D CHO model. The
initial conditions and parameters are set to similate different systems,
as given by Tables \ref{tab:mass-model} and \ref{tab:model}, including
CO/Cu, CH$_4$/Au, H$_2$O, and H + H$_2$. In Figure \ref{fig:phase},
the horizontal axis gives propagation time, while the left and right
vertical axes indicate the phase angle and the time-dependent expectation
of parameter $\eta$,
\begin{equation}
\big\langle\eta\big\rangle(t)=
\frac{\langle\Psi(t)\vert\eta\vert\Psi(t)\rangle}{\langle\Psi(t)\vert\Psi(t)\rangle},
\label{eq:expect-eta-0}
\end{equation}
respectively. For the 3D CHO model, the inner product
$\langle\Psi(t)\vert\Psi(t)\rangle$ remains time-independent because
the Hamiltonian $\hat{H}_{\mathrm{3D-CHO}}$ has no absorbing potential.
In Figure \ref{fig:phase}, the blue symbol $\times$ means the critical point $\eta_c=\sqrt{1+m_1/\max(m_2,m_3)}$
(referred to the right axis), where the phase should change. As shown
in Figure \ref{fig:phase}(a) for CO/Cu, the phase angle changes at
$t\sim178$ fs, while the computed expectation $\langle\eta\rangle$ of
$1.204$ is very close to the critical point of $1.119$. Similarly, as
shown in Figure \ref{fig:phase}(b) for CH$_4$/Au, the phase angle
changes at $\sim950$ fs where the computed $\langle\eta\rangle$ value
of $1.107$ is close to the critical point of $1.037$. The similar
situation can be found in Figure \ref{fig:phase}(c) for H$_2$O, where
the phase angle changes at $\sim48$ fs and computed $\langle\eta\rangle$
of $0.907$ is close to the critical point of $1.031$. Moreover, the
order of critical times perfectly follows the sequence of mass
magnitudes, CH$_4$/Au ($212.967$ dalton and $\sim950$ fs), CO/Cu
($91.546$ dalton and $\sim178$ fs), and H$_2$O ($18.000$ dalton and
$\sim48$ fs). Turning to Figure \ref{fig:phase}(d) for H + H$_2$,
however, the situation changes slightly where the phase angle becomes
linear dependence on time after $\sim10.2$ fs with computed $\langle\eta\rangle$
value of $0.597$ that is largely smaller than the critical point of
$\sqrt{2}$. But, one can still find a change of phase angle from $\sim\pi$
to zero at $\sim10.2$ fs, which indicates that the wave function changes
its sign.

\begin{figure}[h!]
 \centering
  \subfigure[\quad Time-dependent phase of the 3D REX model for H + H$_2$]{
   \includegraphics[width=18cm]{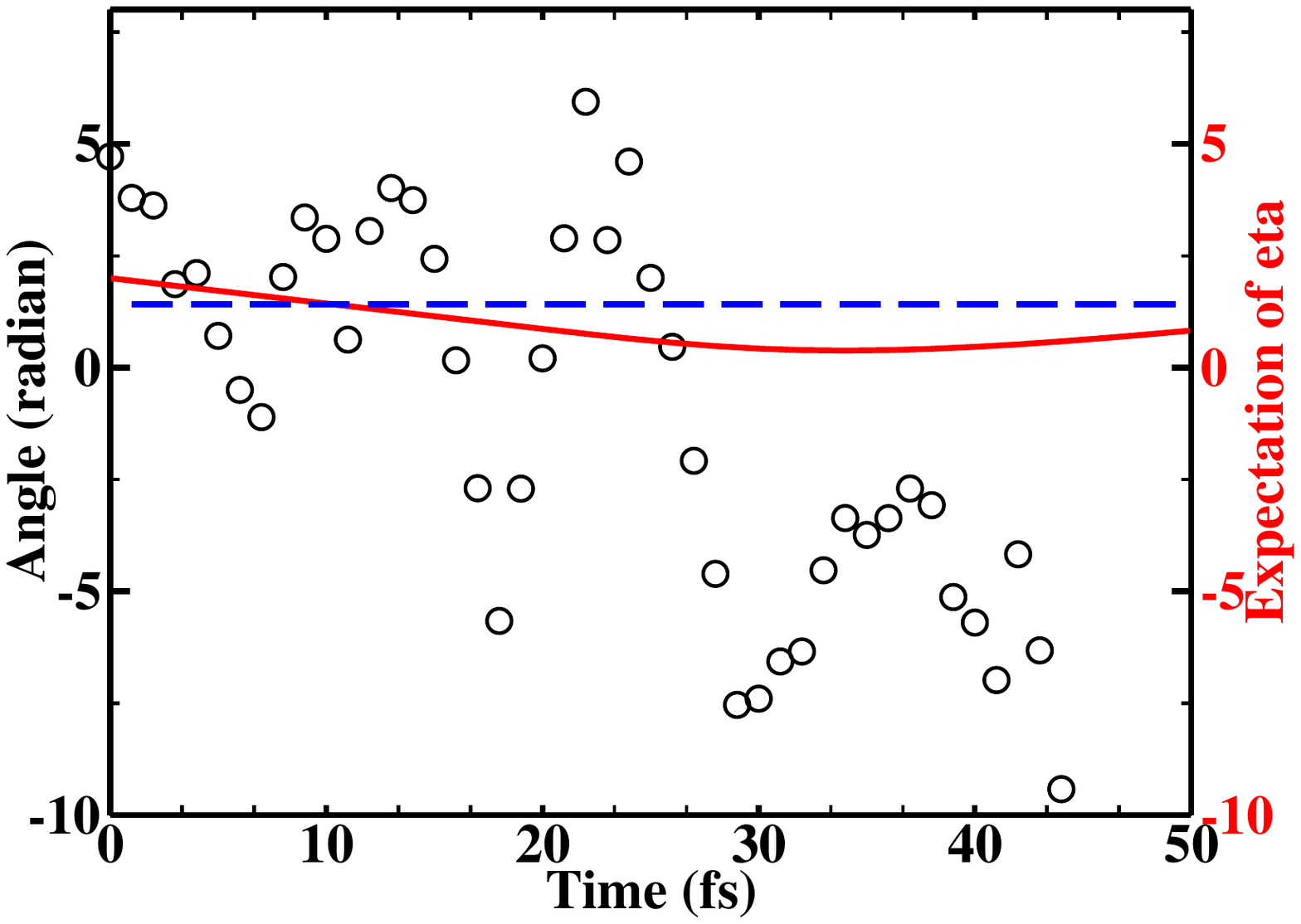}
}  
     \caption{\figfoot}
      \label{fig:phase-prop}
       \end{figure}

Having shown the results of bounded models through the 3D CHO Hamiltonian,
we turn to Figure \ref{fig:phase-prop} for the time-dependent phase angle
computed by the 3D REX model. Similar to Figure \ref{fig:phase}, the
initial conditions and parameters are set to similate different molecular
systems, as given by Tables \ref{tab:mass-model} and \ref{tab:model}.
Results for these systems are illustrated by subfigures (a)-(d) for
CO/Cu, CH$_4$/Au, H$_2$O, and H + H$_2$, respectively. In Figure
\ref{fig:phase-prop}, the time-dependent expectations $\langle\eta\rangle(t)$
are also illustrated by red lines where the inner product of $\Psi(t)$
remains time-independent due to no absorbing potential in the Hamiltonian.
Again, the blue symbol $\times$ means the critical point $\eta_c$.
Moreover, since the phase of H + H$_2$ changes dramatically, one cannot
clearly find transition of the phase angle and thus the time where the
critical point $\sqrt{2}$ occurs. For clarity, the critical point is
given by a blue dashed line in Figure \ref{fig:phase-prop}(d). As shown
in Figure \ref{fig:phase-prop}(a) for CO/Cu, the phase angle changes at
$t\sim266$ fs, while the computed expectation $\langle\eta\rangle$ of
$1.504$ is close to the critical point of $1.119$. The similar situation
can be found in Figure \ref{fig:phase-prop}(c) for H$_2$O, where the
phase angle changes at $\sim34$ fs and computed $\langle\eta\rangle$
of $1.667$ is approximately close to the critical point of $1.031$.
As shown in Figure \ref{fig:phase-prop}(b), the phase angle for CH$_4$/Au
has an almost constant plateau in the region of $238\sim242$ fs. By
Equation \eqref{eq:solution-of-se-000}, although there is no phase
transition, a discontinuous phase change exists if the dynamical phase
is subtracted. Based on the phase reduction rate of $\pi/2$ per fs as
illustrated in Figure \ref{fig:phase-prop}(b), the geometric phase angle
would have undergone a shift of nearly $\pi$ over the time interval
$238\sim242$ fs. At $\sim240$ fs the computed $\langle\eta\rangle$
value of $1.728$ is approximately close to the critical point of
$1.037$. Turning to Figure \ref{fig:phase-prop}(d) for H + H$_2$,
the situation becomes unclear where the phase angle becomes linear
dependence on time after $\sim35$ fs. Below $30$ fs, the phase angle
changes with amplitude of $2\pi$ (at $\sim10$ fs) or $4\pi$ (at
$\sim20$ fs). At $\sim33$ fs, there is a phase angle change of $\pi$
with computed $\langle\eta\rangle$ value of $0.381$ that is largely
smaller than the critical point of $\sqrt{2}$. This indicates that
the wave function changes its sign at $\sim33$ fs but changes phase
angle below $30$ fs. For clarity, the critical point is shown by a
blue dashed line in Figure \ref{fig:phase-prop}(d).

Comparing phase results for CH$_4$/Au and CO/Cu (see Figures
\ref{fig:phase}(a)-(b) and \ref{fig:phase-prop}(a)-(b)) with those for
H$_2$O and H + H$_2$ (see Figures \ref{fig:phase}(c)-(d) and
\ref{fig:phase-prop}(c)-(d)), one can find the following two features.
If $m_3$ is very larger than $m_1$ and $m_2$, as shown by results for
for CO/Cu and CH$_4$/Au, the phase angle just changes by $\pi$ near
the critical point $\eta_c$, maintaining an approximately linear
relationship with time throughout the other propagation duration
(see the black and red lines of Figure \ref{fig:tot-phase}). If
$m_3$ is not very large or even relatively small, as shown by results
for H$_2$O and H + H$_2$, the phase angle remains nearly constant in
the early propagation steps, changes by $\pi$ and then exhibits an
approximately linear relationship with time (see the blue and green
lines of Figure \ref{fig:tot-phase}). The linear relationship between
the phase angle and time can be found by the dynamical phase in Equation
\eqref{eq:solution-of-se-000}. Moreover, for larger difference between
$m_3$ and $\max{m_1,m_2}$, the phase varies more smoothly over time than
that for smaller difference, where the phase changes with more fluctuations
as illustrated in Figure \ref{fig:tot-phase}. Such significant differences
in phenomena should be arisen from the mass differnce between $m_3$
and $\max{m_1,m_2}$. Comparing the 3D REX model with the 3D CHO model,
one can find the following two different points. First, comparing
resulting for CH$_4$/Au ($212.967$ dalton and $\sim240$ fs) with those
for CO/Cu ($91.546$ dalton and $\sim266$ fs), the order of critical
times might not follow the sequence of mass magnitudes. It should be
noted that difference of $26$ fs between $240$ fs and $266$ fs is rather
small and the reaction time based on the unperturbed $V_3$ term is about
$300$ fs, which is very close to the critical times of $240$ fs and $266$
fs. Thus, the mass ordering inversion is not a major concern. Second,
the 3D CHO model describes the linear relationship betwen the phase and
time, together with the phase transition phenomena better than the 3D
REX model. This might be arisen from the fact that the CHO and REX models
focus on bound state and scattering state, respectively. Bound states
typically conserve energy and exhibit linear relationship between phase
and time, whereas scattering states may violate energy conservation
(particularly when an complex absorbing potential is introduced),
showing irregular phase changes.

\begin{figure}[h!]
   \centering
    \includegraphics[width=18cm]{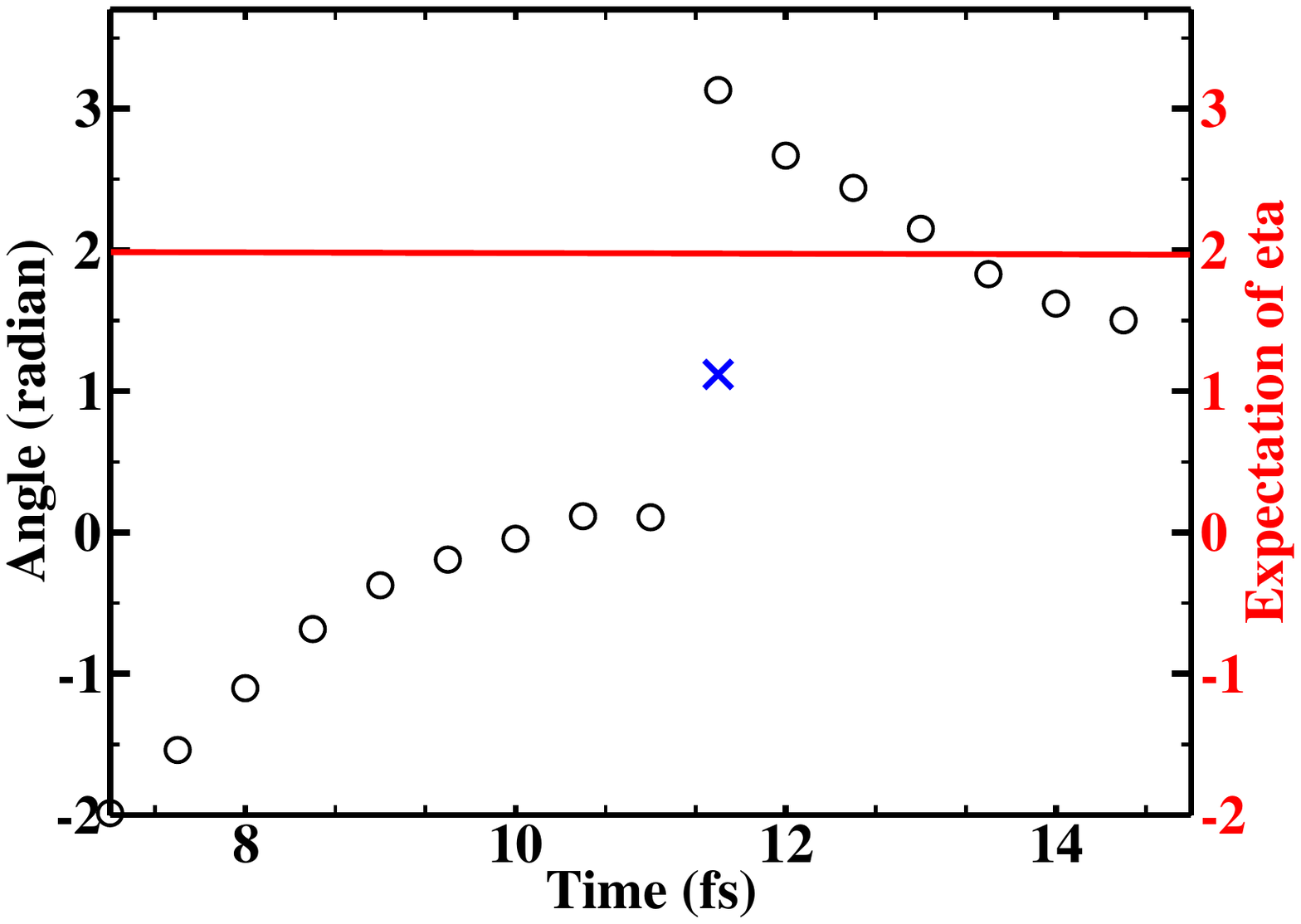}
     \caption{\figfoot}
      \label{fig:98d-co-cu-dyn}
       \end{figure}

Having demonstratd the present theory by the 3D CHO and 3D REX models,
next we should turn to numerical demonstrations on high-dimensional
model, where the atomic arrangement is given in Figure \ref{fig:geom-ben}.
Based on the present 3D CHO model (see Equation \eqref{eq:3d-cho-hamiltonian-00}
and Figure \ref{fig:cho}), a model of a diatomic molecule adsorbed on
a metal surface is proposed, where the masses and frequencies are set
to be $\{m_1,m_2,m_3\}$ and $\{\omega_1,\omega_2,\omega_3\}$, respectively.
For simple, the CO/Cu(100) system is chosen as the benchmark. The Cu(100)
surface is supposed to be a three-layer grid with $30$ atoms with mass
$m_3=63.546$ dalton and coordinates $\{Q_{i}\mu\}_{i=0}^{29}$, where
the atom \#0 is the top site, the atoms \#1-\#8 the first layer, the
atoms \#9-\#20 the second layer, the atoms \#21-\#29 the third layer
(see Figure \ref{fig:geom-ben}). The molecular motions are abstractly
described by coordinates set $\mathbi{q}=\{b,\chi,\phi\}\oplus\{R,\vartheta,z\}$.
The internal coordinates set $\{b,\chi,\phi\}$ describes the C-O stretch
and two-dimensional rotation of CO with polar angle $\chi$ and azimuthal
angle $\phi$, while $\{R,\vartheta,z\}$ describes rotation of the entire
molecule along a loop with radius $R_0$ and angle $\vartheta$. By
Equations \eqref{eq:wf-factorization-000}, the Hamiltonian is given by
\begin{equation}
\hat{H}=\hat{H}_{\mathrm{mol}}+\hat{H}_{\mathrm{surf}}'+\hat{H}_{\mathrm{3D-CHO}}=
\hat{T}_{\mathrm{mol}}+\hat{V}_{\mathrm{mol}}+\hat{T}_{\mathrm{surf}}'
+\hat{V}_{\mathrm{surf}}'+\hat{H}_{\mathrm{3D-CHO}}.
\label{eq:model-ham-000}
\end{equation}
where $\hat{H}_{\mathrm{mol}}$ and $\hat{H}_{\mathrm{surf}}'$ describe
motions of CO and Cu(100), respectively, while $\hat{H}_{\mathrm{3D-CHO}}$
is the coupling term between CO and Cu(100) and given by Equation
\eqref{eq:3d-cho-hamiltonian-00}. The Hamiltonian $\hat{H}_{\mathrm{mol}}$
for CO reads
\begin{align}
\hat{H}_{\mathrm{mol}}={}&\hat{T}_{\mathrm{mol}}+\hat{V}_{\mathrm{mol}}
\allowdisplaybreaks [4] \nonumber \\
={}&-\frac{1}{2m}\left[\frac{1}{b^2}\frac{\partial}{\partial b}
\left(b^2\frac{\partial}{\partial b}\right)+
\frac{1}{b^2\sin\chi}\frac{\partial}{\partial\chi}
\left(\sin\chi\frac{\partial}{\partial\chi}\right)+
\frac{1}{b^2\sin^2\chi}\frac{\partial^2}{\partial\phi^2}\right]
\allowdisplaybreaks [4] \nonumber \\
&-\frac{1}{2(m_1+m_2)}\left(\frac{1}{R}\frac{\partial}{\partial R}
R\frac{\partial}{\partial R}+
\frac{1}{R^2}\frac{\partial^2}{\partial\vartheta^2}\right)+
\frac{1}{2}k_b\big(b-b_0\big)^2+\frac{1}{2}k_R\big(R-R_0\big)^2,
\label{eq:model-ham-001}
\end{align}
where $m=m_1m_2/(m_1+m_2)$ is the resuced mass of CO. The radial potential
$\hat{V}_{\mathrm{mol}}=k_R(R-R_0)^2/2$ ensures that the CO molecule always
move along the closed loop. Motions of the Cu(100) surface are described
by \cite{men21:2702}
\begin{align}
\hat{H}_{\mathrm{surf}}'=-\frac{1}{2m_3}\sum_{\mu=x,y}\frac{\partial^2}{\partial Q_{0\mu}^2}-
\frac{1}{2m_3}\sum_{i=1}^{29}\sum_{\mu=x,y,z}\frac{\partial^2}{\partial Q_{i\mu}^2}
+\sum_{i=0}^{29}\sum_{\mu=x,y,z}V_{\mathrm{M}}\big(Q_{i\mu}-Q_{i\mu}^{(0)}\big)+
\frac{1}{2}\sum_{0\leq i<j\leq29}k_Z\Big(Q_{iz}-Q_{jz}\Big)^2,
\label{eq:model-ham-002}
\end{align}
where $Q_{i\mu}^{(0)}$ represents optimized geometry of the $i$-th surface
atom and $V_{\mathrm{M}}(\rho)$ is a Morse potential with parameters $D$
and $a$,
\begin{equation}
V_{\mathrm{M}}(\rho)=D\big[\exp(-2a\rho)-2\exp(-a\rho)\big]
\label{eq:model-ham-003}
\end{equation}
Such surface model has been used to study surface scattering dynamics
of the CO/Cu(100) system \cite{men21:2702}. One should note that the
3D CHO Hamiltonian $\hat{H}_{\mathrm{3D-CHO}}$ already contains the
Hamiltonian for out-of-plane motion of the top atom and the coupling
term between molecule and surface. Table \ref{tab:98d-model} gives
parameters of the present 98D Hamiltonian. Table \ref{tab:98d-num-prop}
gives numerical details of the present 98D ML-MCTDH calculations, where
Figure \ref{fig:ml-tree-98d} illustrates the multilayer tree-structure
(called ML-tree) of the 98D wave function. We refer the reader to References
\cite{wan03:1289,man08:164116,ven11:044135} for further details on
ML-MCTDH and ML-tree. By the present 98D ML-MCTDH calculations, one
can find the phase shift as shown in Figure \ref{fig:98d-co-cu-dyn}.
This 98D results are similar to those of the low-dimensional models.

\subsection{Discussions on Experiments\label{sec:expt}}

It was found that \cite{wan99:4510,yoo03:9568} the fundamental CH$_3$
symmetric-stretching mode (called $v_1$) and one of the first overtones
of the CH$_3$ degenerate deformation mode (called $v_5$) couple strongly
via a Fermi interaction leading to two eigenstates splited by $\sim59$
cm$^{-1}$. This indicates existence of a pair of Fermi states of the
CH$_3$ symmetric-stretching mode $v_1$, called $v_1$-I and $v_1$-II.
These two eigenstates are linear combinations of two zero-order modes
in symmetric and antisymmetric forms. Two Fermi-coupled vibrations
execute synchronized motions in the $v_1$-$v_5$ space with comparable
amplitudes, but differing in their relative phases. By their
crossed-molecular-beam experiments, Pan and Liu \cite{pan22:545} found
Fermi-phase-induced interference phenomenon in the Cl + CH$_3$D
($v_1$-I/$v_1$-II) $\to$ CH$_2$D ($v=0$) + HCl ($v$) reaction and
proposed nature of this phenomenon. First, this reaction has a late
barrier with a product-like transition state (TS) in the exit channel
which implies that the reactant needs appropriate motions to turn the
corner toward the TS, as expected by Polanyi's rule \cite{pol87:952,pol87:680}.
Classically, for the in-phase $v_1$-I reagent, the CH$_3$D molecule
first stretches and bends, then shrinks and straightens. In contrast,
for the out-of-phase $v_1$-II, it is most likely to be found the
CH$_3$D molecule either stretched and near planar, or squeezed and
bent. Due to the multidimensional nature of the system, the corner-turning
effects are of multiple folds. As CH$_3$D hits the corner-turning wall,
the impulse may either act in concert or counteract the incoming Fermi-coupled
vibrational motions. By such analysis based on classical-mechanical view,
Pan and Liu \cite{pan22:545} proposed that the prime factor to consider
is how well the force vectors along the stretching and bending directions
match with the in-phase or out-of-phase oscillation to allow the appropriate
pathway towards the TS. By the present theory on the adiabatic and
diabatic EOMs of Equations \eqref{eq:new-adi-dia-rep-016} and \eqref{eq:new-adi-dia-rep-017},
it should be possible to interpret the above experiments by solve an
appropriate EOM on coupled-vibrational system. For instance, separating
strongly coupled modes, say $v_1$ and $v_5$, from the other vibrational
modes, one can construct the Hamiltonian model in Equation \eqref{eq:wf-factorization-000}.
Then, letting the $v_1$-I and $v_1$-II states be basis in Equation
\eqref{eq:adi-dia-rep-002}, one can obtain the EOMs of the other modes.
By solving such matrix EOMs, we can explore how the $v_1$-I and $v_1$-II
states influence the reaction dynamics of Cl + CH$_3$D ($v_1$-I/$v_1$-II).
Such dynamics is non-adiabatic at the level of nuclear motions rather
than that of electronic motions. Of course, the core questions of the
above calculations are how to construct the Hamiltonian model and which
EOM is adopted. These two questions are similar to those on electronic
structure in non-adiabatic chemical dynamics.

Next, Beck and co-workers \cite{rei25:962} reported several results by
their experiments on CH$_4$/Au(111). First of all, Beck and co-workers
\cite{rei25:962} found the implications of the strict magnetic quantum
number conservation required for rovibrational state interference (RSI)
channel suppression. Such state purity in the RSI-allowed levels was
unobserved in molecule-surface scattering before this recently reported
experiments. Moreover, rotation in the plane parallel to the surface
must be strongly decoupled from the decohering influence of the surface
phonons \cite{rei25:962}. Obviously, this is consistent with previous
theoretical calculation results obtained from the lattice effects on
the adsorption dynamics of CO/Cu(100) \cite{men13:164709,men15:164310,men17:184305,men21:2702}. 
Lastly, these experimental results \cite{rei25:962} serve as a particularly
cogent illustration of the role of symmetry. To explain the striking
regularities in the experiments, Beck and co-workers \cite{rei25:962}
considered the existence of a continuous family of reflection symmetries
in the polyatomic-molecule-surface interaction, where all of these
symmetries are discrete in that their actions always in a finite change
in the molecular coordinates. Due to the inapplicability of the N{\"o}ther
theorem, discrete symmetries are confined to a limited role in a
classical-mechanical context. However, the quantum-mechanical
superposition elevates the role of reflections and other discrete
symmetries. The high contrast interference phenomena \cite{rei25:962}
demonstrate that quantum effects are dominate even for complex
molecular-surface scattering at room temperature.

\begin{figure}[h!]
   \centering
    \includegraphics[width=18cm]{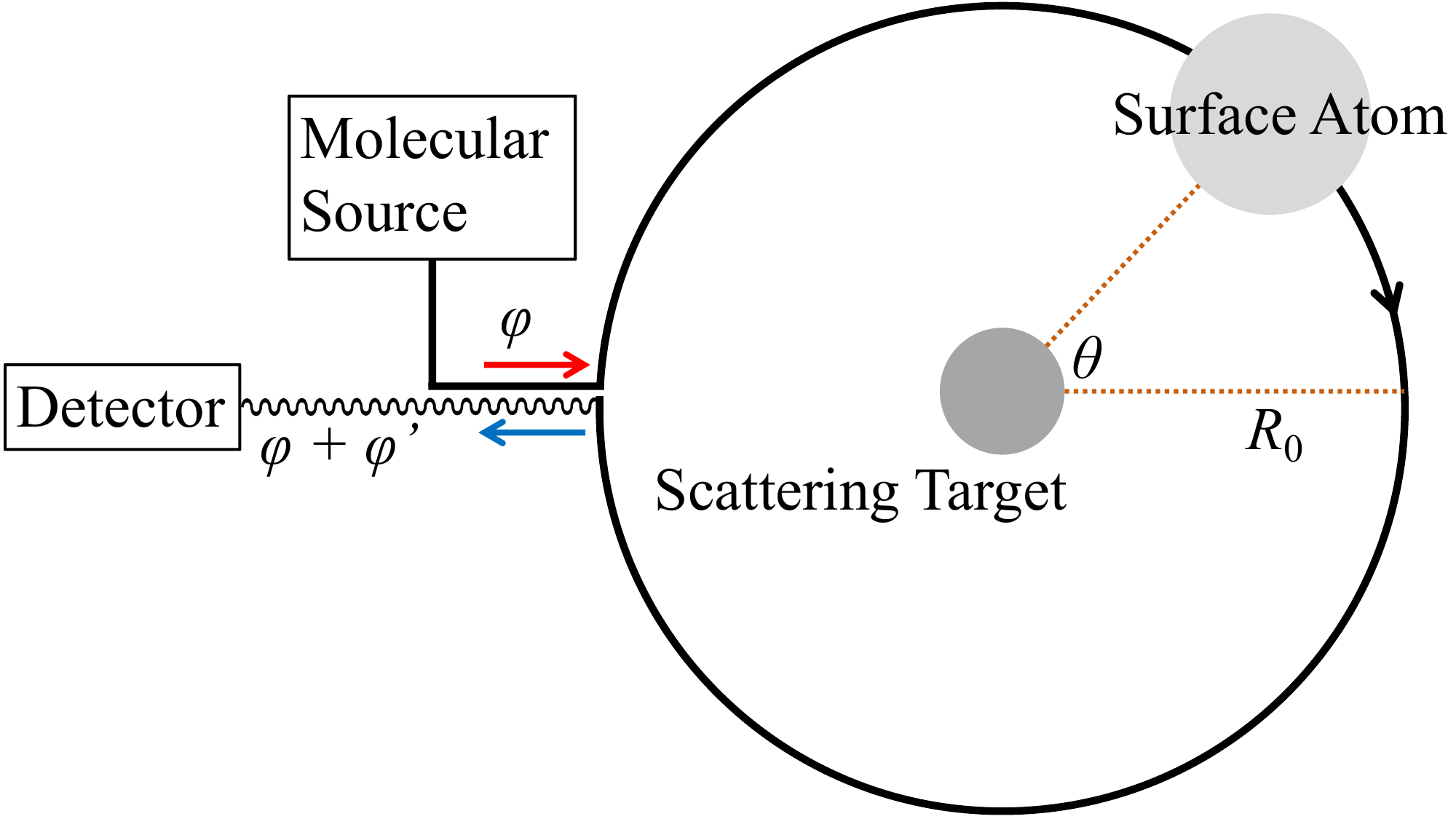}
     \caption{\figfoot}
      \label{fig:expert}
       \end{figure}

Drawing inspiration from experiments reported by Yang and co-workers
\cite{xie20:767} and Beck and co-workers \cite{rei25:962}, an ideal
experiment as illustrated by Figure \ref{fig:expert} is proposed to
verify the sign change of the molecular wave function as indicated by
Equation \eqref{eq:adi-dia-rep-014}. To understand the quantum
interference effects, Yang and co-workers \cite{xie20:767} detected
unusual oscillations in the differential cross section by measuring
two topologically distinct pathways in the H + HD $\to$ H$_2$ + D
reaction. The notable oscillation patterns are attributed to the strong
quantum interference between the direct abstraction pathway and an
unusual roaming insertion pathway. Such interference pattern \cite{xie20:767}
also provides a sensitive probe of the geometric phase effect at an
energy far below the conical intersection, clearly demonstrating the
quantum nature of this reaction. The ideal experiment (see also Figure
\ref{fig:expert}) can be carried out using a molecule-beam apparatus
based on light molecule, say hydrogen or methane, which scatters from
a metal surface, say gold, held at given surface temperature. A pump
laser might be adopted for selecting rovibrational states of the moelcule.
The surface should be rotatable around the center of the scattering
target region, so that possible to vary the incident angle $\theta_{\mathrm{ini}}$.
The scattered molecule was detected by a tagging laser combined with a
bolometer detector to permit state–resolved population measurements
of the scattered molecule. The incident angle and scattering angle
$\theta_{\mathrm{fin}}$, defined with respect to the surface normal,
are independently variable. Time-of-flight (TOF) spectra at different
incident angle may be further accumulated at the corresponding backward
scattering direction. As illustrated by Figure \ref{fig:expert}, given
the incident molecular wave function $\varphi$, the scattered and
incident molecular wave functions superimpose to $\varphi+\varphi'$.
Here, $\varphi'$ means scattered wave function that depends on
rotational coordinate of the surface atom $\theta$. In this process,
both $\theta_{\mathrm{ini}}$ and $\theta_{\mathrm{fin}}$ are
determined by rotation radius $R_0$ of the surface atom, which is
represented by $\theta$. Because of the phase shown by Equation
\eqref{eq:adi-dia-rep-014}, if there is no reaction and other
transition one may observe that the intensity of molecular
signal in the form
\begin{equation}
\vert\varphi+\varphi'\vert^2=2\vert\varphi\vert^2\big(1+\cos\vartheta\big).
\label{eq:singal-00}
\end{equation}
Since $\varphi+\varphi'$ which should depend on $\theta$, by
Equation \eqref{eq:singal-00} one can connect the phase $\vartheta$
with experimental angle $\theta$.

\subsection{Geometric Perspectives\label{sec:geom-inter}}

The Berry phase effects on electronic properties \cite{xia10:1959} play
a fundamental role in describing the geometric features as the parameters
are varied. Similarly, the present work finds the Berry phase effects on
nuclear properties. Now, it is necessary to discuss this issue from the
view of geometry. To this end, we should turn to the concept of Berry
connection
$A_n(\mathfi{z})=\mi\langle\psi_n\vert\partial_{\mathfi{z}}\vert\psi_n\rangle_{\mathbi{q}}$
(see Equation \eqref{eq:solution-of-se-001}). One of examples is the
NACM term in the adiabatic EOMs (see Equations \eqref{eq:adi-dia-rep-001}
and \eqref{eq:new-adi-dia-rep-016}). In differential geometry, the
concept of connection is introduced to compare vectors in nearby vector
spaces enabling us to find whether the direction of a vector is changed
when it parallel transports along a path (for example, the close path $\Gamma$ in
Equation \eqref{eq:adi-dia-rep-010}). The parallel transport does not
change length and direction of the tangent vector of a geodesic, which
can be seen as a ``straight line'' in curved manifolds. For example,
in flat Euclidean space, a vector undergoing parallel transport along
a straight line, {\it i.e.} geodesic, does not change its length or
direction. Extending to flat Hilbert space, the adiabatic motion of the
quantum state in it is a typical example of the parallel transport, where
the norm and phase of the state remain invariant. In this case, the
motion is unaffected by the slow variation of external parameters. Now,
introducing the parameter space the flat space might become curved due
to the changing external parameters. In this case, answering the question
whether the direction of a vector is changed when it parallel transports
along a close path can be used to figure out whether the space is curved.
For example, parallel transport of a vector along a M{\"o}bius band
results in a reversal of the vector orientation because the non-trivial
monodromy of the band connection induces a $\pi$-rotation under parallel
transport. Similar analysis on the Klein bottle can be given indicating
parallel transport of a vector in the curved space. Its non-abelian holonomy
could constrain wave function phase choices in low-dimensional systems.
Turning to Equation \eqref{eq:solution-of-se-000}, if the space is curved
(by, for example, introducing parameter space as in the present theory),
the quantum state is parallel transported along the curve path making
its phase change. This leads to the Berry phase whose correspondence
in geometry is the holonomy angle. In general, the holonomy angle
quantifies the Berry phase accumulated by the wave function after
parallel transport in parameter space. For a closed loop, this angle
is determined by the integral of the Berry connection, as shown by
Equations \eqref{eq:solution-of-se-001} and \eqref{eq:geome-phase-0000}.

As shown in Section \ref{sec:results-num}, the phase angle is changed
by $\pi$ at the critical point in the parameter space. Now, we should
turn to the question where the phase angle will be changed. In the
theory beyond the BOA, the electronic wave function changes its sign
by adiabatically moving the molecular system ({\it i.e.} parallel
transport of its quantum state) along the close path in the $g$-$h$
space around one CI point (see Section \ref{sec:phase-shift}). If the
close path encircles two CI points, however, the electronic wave
function will not change its sign. This topological characteristic is
usually quantified by a topological invariant, called Chern number,
which characterizes certain properties of phases in quantum systems.
With the Berry connection $A_n(\mathfi{z})$ in Equation \eqref{eq:solution-of-se-001},
the Chern number is defined as
\begin{equation}
C_n\propto\int_{\Gamma}\mathrm{d}\mathfi{z}\Big(
\partial_{\mathfi{z}}\times A_n\big(\mathfi{z}\big)\Big)
=\int_{\Gamma}\mathrm{d}\mathfi{z}F_n\big(\mathfi{z}\big),
\label{eq:chern-number}
\end{equation}
where $F_n(\mathfi{z})=\partial_{\mathfi{z}}\times A_n(\mathfi{z})$ is
the Berry curvature for the $n$-th state. The integral of Equation
\eqref{eq:chern-number} is taken over the parameter space along the
path $\Gamma$. According to Equation \eqref{eq:chern-number}, the Chern
number counts the number of times the Berry curvature ``wraps'' around
the parameter space and the number of the crtical point. The latter
gives the number of changing the sign of the wave function. Taking
the M{\"o}bius band again as an example, an odd Chern number implies
a M{\"o}bius strip-like twist in the bundle, causing the wave function
to flip sign after traversing a loop. Therefore, if the Chern number is
odd, the wave function changes sign; if even, it does not. In the present
theory for nuclear evolution, the parameter space contains enviramental
conditions which vary slowly leading to the adiabatic evolution of the
molecular system. The Chern number counts the number of the critical
points, where the conditions of adiabatic evolution are broken.
In Table \ref{tab:geom-comp}, we collect the above geometric features
of the Berry phase effects for both electron and nuclear, together with
their relationships with the corresponding electromagnetic features,
demonstrating that the critical point geometrically exhibits characteristics
similar to those of charges. Therefore, the separation of the DOFs of
the molecular system also leads to fruitful geometric characteristics,
like those in electronic-structure theory, condensed matter physics,
and field theory.

\section{Conclusions\label{sec:con}}

In this work, the Berry phase effects of the nuclear wave function
are found and discussed by considering the essence in separating
reactant degrees of freedom (DOFs). The phase effects on electronic
properties give rise to quantum interferences already observed by
experiments \cite{che21:936,rei25:962}. In quantum dynamics calculations on high-dimensional
reaction, dimensionality reduction is often employed to save computational
cost. For instance, in constructing the PES of molecule-surface system
and then propagating its wave function, the surface atoms are usually
supposed to be fixed at their optimal conformation. In calculations on quantum state-resolved dynamics of a chemical
reaction, reactants are usually prepared in separated eigenstates of
individual fragments, and their direct-product is then evolved in time.
In this work, we focus on the essence in separating them and the Berry
phase effects of the nuclear wave function. 
By the present theory, mechanism
of inter/intramolecular energy redistribution is developed to deeply
understand reactive dynamics with multirovibrational states. To
demonstrate the present theory, two three-dimensional (3D) models
reductively describing the molecular reaction are developed to
simulate transport of the system along a closed
loop in a parameter space represented of inter/intramolecular energy
transfers. Employing these models, extensive multiconfigurational
time-dependent Hartree (MCTDH) calculations are performed to solve the
time-dependent nuclear Schr{\"o}dinger equation at various initial
conditions. These calculations clearly indicate quantum interference
in the parameter space. Discussions on the separation of the reactants
are made, while perspectives on the Berry phase effects predicted by
the present work are given from the viewpoint of differential geometry.
As a conclusion remark, the Berry phase effects on molecular dynamics
are also thoroughly compared with those on electronic properties and
mode/bond-specific reactivity.

%

\section*{Acknowledgements}

The financial supports of National Natural Science Foundation of China
(Grant No. 22273074) and Fundamental Research Funds for the Central
Universities (Grant No. 2025JGZY34) are gratefully acknowledged. The
authors also thank the {\it Centre National de la Recherche Scientifique}
(CNRS) International Research Network (IRN) ``MCTDH'' which is a
constructive platform for discussing quantum dynamics. The authors are
grateful to Dr. C. S. Reilly (EPFL) and Prof. Dr. H.-D. Meyer (Heidelberg)
for helpful discussions and also grateful to anonymous reviewers for their
thoughtful suggestions.

\clearpage
\begin{sidewaystable}
\caption{%
Comparison of the present separation theory in molecular scattering
(MS) with that in electron-nuclear (EN) system by several criteria
(given in the second column). Both separation theories focus on
dynamics of fast sub-systems, {\it i.e.}, motions along reaction
coordinate and electronic motions in chemical reaction. The third
and fourth columns give the characteristics of the MS and EN separations,
respectively. The rightmost column gives remarks on corresponding
comparison.
}%
 \begin{tabular}{lclclclcr}
  \hline
No. &~~& Criteria &~& MS &~& EN &~~& Remark  \\
\hline
1 && slow object && inactive fragments && nuclear  && heavy part in the system
\footnote{According to Equations \eqref{eq:wf-factorization-011} and
\eqref{eq:adi-dia-rep-001}, eigen-energy values of slow and fast
objects are $\mu_1$ and $E(\mathbi{Q})$, respectively.
\label{foot-energy-slow-fast}}  \\
2 && fast object && active fragments && electron && light part in the system
\textsuperscript{\ref{foot-energy-slow-fast}} \\
3 && degeneracy space && parameter space && branching $g$-$h$ space
\footnote{The gradient difference $g$ describes the gradient
difference between the two electronic states, while the non-adiabatic
coupling $h$ describes the strength of coupling.}
  && the degeneracy is lifted linearly \\
4 && degeneracy object  && motion and assignment && electronic energy 
&& physical quantities degenerated \\
&& && && && in the degeneracy space   \\
5 && degeneracy topology && semi-conical intersection
\footnote{The 2D CHO model predicts a semi-conical intersection topology
as previously discussed \cite{men15:164310,men17:184305}.}
&& conical intersection && topology of the degeneracy  \\
 && && && && object in the degeneracy space \\
6 && phase       && molecule
\footnote{If and only if the system moves along a closed loop in the
parameter space, the wave function of the light part undergoes a phase
change. \label{foot:phase}}
&& electron \textsuperscript{\ref{foot:phase}}
&& see Equations \eqref{eq:effective-ham-00} and \eqref{eq:adi-dia-rep-014} \\
7 && coupling && mass coupling
\footnote{Coupling between the molecule and surface is arisen from mass
ratio between the molecule and surface.}
&& electromagnetic coupling \footnote{The Born-Oppenheimer
approximation is arisen from electronic motions in the molecular
system} && in essence, mass coupling \\
8 && adiabatic dynamics && molecular phase changed && electronic phase changed &&
invariant state of the fast object  \\
9 && diabatic dynamics && IMR/IVR process && electronic-state transition &&
multi-state transition \\
 && && && && of the fast object  \\
10 && non-adiabatic dynamics && IMR/IVR process && electronic-state transition &&
model coupling among \\
 && && && &&  states of the fast object  \\
\hline
\end{tabular}
\label{tab:compare-phase}
\end{sidewaystable}

\clearpage
\begin{sidewaystable}
 \caption{%
Masses and parameters of the models used in the quantum dynamics
calculations for demonstrating the present theory. The second, third,
and fourth columns give masses (in dalton) of $m_1$, $m_2$, and $m_3$,
respectively. The fifth column gives the imitation of model, including
H + H$_2$, H + OH, CO/Cu(100), and CH$_4$/Au(111). The sixth and seventh
columns give values of $\zeta_1$ and $\zeta_2$ in the 3D CHO model
(see Equation \eqref{eq:3d-cho-hamiltonian-00}), respectively. The
eighth and ninth columns give values of $\kappa_1$ and $\kappa_2$ in
the 3D REX model (see Equation \eqref{eq:reax-3d-rex-0001}), respectively.
The rightmost column gives remarks of the model.
}%
 \begin{tabular}{lcccccccccccccccccr}
  \hline
No. &~~& \multicolumn{7}{c}{Mass models} &~~& \multicolumn{7}{c}{Parameters} &~~& Remark  \\
\cline{3-9} \cline{11-17}
    &~~& $m_1$ &~~& $m_2$ &~~& $m_3$ &~~& Imitation &~~&
    $\zeta_1$ &~~& $\zeta_2$ &~~& $\kappa_1$ &~~& $\kappa_2$ &~~&      \\
\hline
1 && $1.000$ && $1.000$  && $1.000$ && H + H$_2$ && $1.000$ && $1.000$ && $0.125$ && $0.125$ 
&& the basic model  \\
2 && $1.000$ && $16.000$ && $1.000$ && H + OH    && $0.250$ && $4.000$ && $3.257\times10^{-3}$ 
&& $0.443$ && dissociation of water \\
3 && $16.000$&& $12.000$ && $63.546$&& CO/Cu(100)&& $1.155$ && $0.435$ && $4.852$ && $0.322$
&& CO adsorbed on copper \\
4 && $15.000$&&  $1.000$ &&$196.967$&& CH$_4$/Au(111)
&& $3.873$ && $7.125\times10^{-2}$ && $0.928$ && $3.557\times10^{-4}$ 
&& dissociative chemisorption \\
  &&         &&          &&         &&           &&  && && && && of CH$_4$ on gold \\
\hline
\end{tabular}
\label{tab:mass-model}
\end{sidewaystable}

\clearpage
\begin{table}
 \caption{%
Parameters in the present Hamiltonian models based on which the MCTDH
calculations are performed to illustrate the wave function phase. The
upper panel gives the model for the molecule-surface system, while the
lower panel gives the model for the chemical reaction. The second column
gives the parameters in defining the present demonstration models. The
third column gives values of the parameters in the present calculations.
The frequency parameters $\eta$ and $\xi$ of the 3D CHO model represent
states of the reactants, such as the nozzle and target temperatures.
The frequency parameters $\alpha$ and $\beta$ of the 3D REX model represent
internal motions of the reactant fragments revised by the H + H$_2$ system.
The $V_3$ term in $V_{\mathrm{3D-REX}}$ is adopted by the BKMP2 PES of
the H + H$_2$ system \cite{boo96:7139}, while the $V_{\mathrm{rev}}$
term is a perturbation of $V_3$. As usual, the atomic units are used.
The fourth column gives expression in which the parameter appear. The
rightmost column gives remarks on these parameters.
}%
 \begin{tabular}{lclclclcr}
  \hline
No. &~& Parameter  &~& Value &~& Expression &~& Remark  \\
\hline
\multicolumn{9}{l}{{\it Parameters of the Hamiltonian model for surface scattering}}  \\
1 && $m_1$ && ---
\footnote{See Table \ref{tab:mass-model} for various models.\label{foot:models}} 
&& Equation \eqref{eq:3d-cho-hamiltonian-00} 
&& mass of the atom in molecule \\
2 && $m_2$ && --- \textsuperscript{\ref{foot:models}} && Equation \eqref{eq:3d-cho-hamiltonian-00}
&& mass of the atom in molecule \\
3 && $m_3$ && --- \textsuperscript{\ref{foot:models}} && Equation \eqref{eq:3d-cho-hamiltonian-00} 
&&  mass of surface atom \\
4 && $\omega_2$ && $500.0\;\mathrm{cm}^{-1}$
\footnote{It is a typical frequency value for restricted vibration of
the entire adsorbed molecule.}
&& Equation \eqref{eq:3d-cho-hamiltonian-00} && force constant of the adsorption \\
5 && $\eta$ && $\omega_1^2/\omega_2^2$ && Equation \eqref{eq:3d-cho-hamiltonian-00}
&& function of nozzle temperature \\
6 && $\xi$ && $\omega_3^2/\omega_2^2$ && Equation \eqref{eq:3d-cho-hamiltonian-00}
&& function of target temperature \\
\hline
\multicolumn{9}{l}{{\it Parameters of the Hamiltonian model for molecular reaction}}  \\
1 && $m_1$ && --- \textsuperscript{\ref{foot:models}} && Equation \eqref{eq:reax-3d-rex-0001}
&& mass of the fragment $\#1$, the colliding group  \\
2 && $m_2$ && --- \textsuperscript{\ref{foot:models}} && Equation \eqref{eq:reax-3d-rex-0001}
&& mass of the fragment $\#2$ in the target group  \\
3 && $m_3$ && --- \textsuperscript{\ref{foot:models}} && Equation \eqref{eq:reax-3d-rex-0001}
&& mass of the fragment $\#3$ in the target group \\
4 && $\omega_{23}$ && $1.0\;\mathrm{cm}^{-1}$
\footnote{The value should ensure that the additional term $V_{\mathrm{rev}}$
of Equation \eqref{eq:reax-3d-rex-0000} is small enough and thus
a perturbation.}
&& Equation \eqref{eq:reax-3d-rex-0002}
&& force constant of target bonding  \\
5 && $\alpha$ && $\omega_{12}^2/\omega_{23}^2$ && Equation \eqref{eq:reax-3d-rex-0002}
&& internal motions of fragments $\#1$ and $\#2$ \\ 
6 && $\beta$ && $\omega_{13}^2/\omega_{23}^2$ && Equation \eqref{eq:reax-3d-rex-0002}
&& internal motions of fragments $\#1$ and $\#3$  \\
\hline
\end{tabular}
\label{tab:model}
\end{table}

\clearpage
\begin{sidewaystable}
 \caption{%
DVR-grids used in the present quantum dynamics calculations, and
parameters of the initial guess function for preparing the initial
wave function $\Psi(t=0)$. The definitions of the surface geometry
and coordinates (indicated in the first column) are given in Figure
\ref{fig:cho}. The second
column describes the primitive basis functions, which underlay the DVR.
The third column gives the number of the grid points. The fourth column
gives the range of the grids in atomic unit or radian. The fifth column
gives the symbol of the one-dimensional (1D) function for each coordinate
of the guess function. The initial wave function is then computed by
relexing the guess function. The other columns give the parameters for
these 1D functions, including positions and momenta in the 1D function,
frequency ($\omega_{\mathrm{HO}}$) and mass ($M_{\mathrm{HO}}$) of the
harmonic oscillator (HO) function, width of the Gaussian function ({\it
i.e.}, and variance of the modulus-square of the Gaussian function,
$W_{\mathrm{GAUSS}}$).
}%
 \begin{tabular}{llllllrlrlrlrlr}
  \hline
Coordinates &~~& \multicolumn{5}{c}{Primitive basis function}  
&~~~~~~& \multicolumn{7}{c}{Initial guess function}  \\ \cline{3-7} \cline{9-15}
                  &~~& Symbol 
\footnote{Symbols HO and SIN stand the harmonic oscillator and the sine
DVR basis functions, respectively, while LEG denotes a one-dimensional
Legendre DVR basis.}
                  &~~& Grid points &~~& Range of the grids &~~~~~~& Symbol
\footnote{Symbols HO and GAUSS designate the harmonic oscillator
eigenfunction and Gaussian function, respectively, while LEG
denotes the Legendre function.}
&~~& Position &~~& Momentum &~~& Parameters  \\
\hline
\multicolumn{15}{l}{{\it Coordinates of the molecule-surface model}}  \\
$q_1$  && HO && $15$ && $[-0.800,0.800]$ && HO && $0.0$  && $0.0$  &&
$500.0$ cm$^{-1}$, $m_1$  \\
$q_2$ && HO && $15$ && $[-0.800,0.800]$ && HO && $0.0$  && $0.0$
&& $500.0$ cm$^{-1}$, $m_2$  \\
$q_3$  && HO && $15$ && $[-0.800,0.800]$ && HO && $0.0$  && $0.0$  &&
$500.0$ cm$^{-1}$, $m_3$ \\
$\eta$ && SIN && $68$ && $[0.00,2.50]$
&& GAUSS && $1.5$  && $-15.0$ && $W_{\mathrm{GAUSS}}=0.12$ \\
$\xi$ && SIN && $68$ && $[0.00,2.50]$
&& GAUSS && $1.5$  && $-15.0$ && $W_{\mathrm{GAUSS}}=0.12$ \\
\hline
\multicolumn{15}{l}{{\it Coordinates of the reaction model}}  \\
$r_{\mathrm{d}}$ && SIN && $68$ && $[1.00,9.04]$ && GAUSS && $4.5$ && $-8.0$ &&
$W_{\mathrm{GAUSS}}=0.25$  \\
$r_{\mathrm{v}}$ && SIN && $48$ && $[0.60,6.24]$ && EIGENF
\footnote{Symbol EIGENF means the eigenfunction of a specified potential
which, in this work, is the potential energy of the H$_2$ molecule.}
&& --- && --- && ground state  \\
$\theta$ && LEG && $31$ && $[0,\pi]$ && LEG && --- && --- && $j_{\mathrm{ini}}=0$ \\
$\alpha$ && SIN && $68$ && $[0.00,2.50]$
&& GAUSS && $1.5$  && $-15.0$ && $W_{\mathrm{GAUSS}}=0.12$ \\
$\beta$ && SIN && $68$ && $[0.00,2.50]$
&& GAUSS && $1.5$  && $-15.0$ && $W_{\mathrm{GAUSS}}=0.12$ \\
\hline
 \end{tabular}
   \label{tab:num-prop}
    \end{sidewaystable}

\clearpage
\begin{table}
 \caption{%
Same as Table \ref{tab:model} except for parameters in the present 98D
Hamiltonian model for the molecule-surface system. Parameters of
$\hat{H}_{\mathrm{surf}}'$ given in Equations \eqref{eq:model-ham-002}
and \eqref{eq:model-ham-003} were previously optimzed for the Cu(100)
surface \cite{men21:2702}.
}%
 \begin{tabular}{lclclclcr}
  \hline
No. &~& Parameter  &~& Value &~& Expression &~& Remark  \\
\hline
1 && $m_1$ && $16.000$ dalton && Equation \eqref{eq:3d-cho-hamiltonian-00} 
&& mass of the atom in molecule \\
2 && $m_2$ && $12.000$ dalton && Equation \eqref{eq:3d-cho-hamiltonian-00}
&& mass of the atom in molecule \\
3 && $m_3$ && $63.546$ dalton && Equation \eqref{eq:3d-cho-hamiltonian-00} 
&&  mass of surface atom \\
4 && $\omega_2$ && $500.0\;\mathrm{cm}^{-1}$
\footnote{It is a typical frequency value for restricted vibration of
the entire adsorbed molecule or bonding stretch.\label{foot:vib-frquen}}
&& Equation \eqref{eq:3d-cho-hamiltonian-00} && force constant of the adsorption \\
5 && $\eta$ && $\omega_1^2/\omega_2^2$ && Equation \eqref{eq:3d-cho-hamiltonian-00}
&& function of nozzle temperature \\
6 && $\xi$ && $\omega_3^2/\omega_2^2$ && Equation \eqref{eq:3d-cho-hamiltonian-00}
&& function of surface temperature \\
7 && $k_b$    && $1000.0$ cm$^{-1}$ \textsuperscript{\ref{foot:vib-frquen}}
&& Equation \eqref{eq:model-ham-001} && force constant of the bond \\
8 && $k_R$    && $1.000\times10^{-2}$ hartree 
&& Equation \eqref{eq:model-ham-001} && force constant along the $R$ coordinate \\
9 && $b_0$    &&  $2.267$ bohr && Equation \eqref{eq:model-ham-001} && bond length \\
10&& $R_0$    &&  $0.397$ bohr && Equation \eqref{eq:model-ham-001} && radius of the closed loop \\
11&& $\{Q_{i\mu}^{(0)}\}_{i=0}^{29}$, $\mu=x,y,z$ && See Figure \ref{fig:geom-ben} 
&& Equation \eqref{eq:model-ham-002} && optimized coordinates of the $i$-th atom \\
12 && $d$ && $4.830$ bohr && Equation \eqref{eq:model-ham-002} && lattice constant \\
13 && $k_Z$ && $1.100\times10^{-2}$ hartree && Equation \eqref{eq:model-ham-002}
&& coupling strength the non-top atoms \\
14&& $D$ && $0.300$ hartree && Equation \eqref{eq:model-ham-003} && deep of the Morse potential \\
15&& $a$ && $0.390$ {\AA}$^{-1}$ && Equation \eqref{eq:model-ham-003} 
&& width parameter of Morse well \\
\hline
\end{tabular}
\label{tab:98d-model}
\end{table}

\clearpage
\begin{sidewaystable}
 \caption{%
Same as Table \ref{tab:num-prop} except for the present 98D Hamiltonian
model for the molecule-surface system. The atomic arrangement is shown
by Figure \ref{fig:geom-ben}. Parameters of the 98D Hamiltonian model
is given in Table \ref{tab:98d-model}. The sizes of the DVR basis functions
are also given in Figure \ref{fig:ml-tree-98d}.
}%
 \begin{tabular}{llllllrlrlrlrlrlr}
  \hline
Coordinates &~~& \multicolumn{7}{c}{Primitive basis function}
&~~~~~~& \multicolumn{7}{c}{Initial guess function}  \\ \cline{3-9} \cline{11-17}
                  &~~& Symbol 
\footnote{Symbols HO and SIN stand the harmonic oscillator and the sine
DVR basis functions, respectively, while LEG denotes a one-dimensional
Legendre DVR basis.}
                  &~~& Grid points &~~& Position 
\footnote{Positions of the surface atoms, $\{Q_{i\mu}^{(0)}\}_{i=0}^{29}$,
$\mu=x,y,z$, are illustrated in Figure \ref{fig:geom-ben} and can be
computed by paramter $d=4.830$ bohr, as given in Table \ref{tab:98d-model}.\label{foot:pos-atoms}}
&~~& Parameters &~~~~~~& Symbol
\footnote{Symbols HO and GAUSS designate the harmonic oscillator
eigenfunction and Gaussian function, respectively, while LEG
denotes the Legendre function.}
&~~& Position \textsuperscript{\ref{foot:pos-atoms}}
&~~& Momentum &~~& Parameters  \\
\hline
$b$   && HO && $15$ && $2.267$ bohr && $1000.0$ cm$^{-1}$, $6.857$ dalton
&& HO && $2.267$ bohr  && $0.0$  && $1000.0$ cm$^{-1}$, $6.857$ dalton \\
$\chi$ && PLEG
\footnote{Symbols PLEG and EXP denote two-dimensional Legendre
and exponential DVR basis functions, respectively.\label{foot:pleg}}
&& $15$ && ---  && $\chi\in[0,\pi]$ && KLEG 
\footnote{Symbols KLEG and K denote associated Legendre polynomial and body-fixed magnetic
quantum number, respectively.\label{foot:kleg}}
&& $0.0$ && $0.0$ && nosym
\footnote{It means that no symmetry is set, that is all angle quantum number values are used.} \\
$\phi$ && EXP \textsuperscript{\ref{foot:pleg}}
&& $15$ && --- && $\phi\in[0,2\pi]$ && K \textsuperscript{\ref{foot:kleg}}
&& $0.0$ && $0.0$ && $0$
\footnote{It means that body-fixed magnetic quantum number of initial wave function is zero.}  \\
$z$  && HO && $15$ && $2.835$ bohr && $500.0$ cm$^{-1}$, $28.0$ dalton
&& HO && $2.835$ bohr && $0.0$  && $500.0$ cm$^{-1}$, $1.0$ dalton \\
$R$  && HO && $15$  && $0.397$ bohr && $10^{-2}$ hartree, $28.0$ dalton
&& HO && $0.397$ bohr && $0.0$ && $10^{-2}$ hartree, $28.0$ dalton  \\
$\vartheta$ && LEG  && $9$ && --- && $\varphi\in[0,2\pi]$ 
&& GAUSS && $0.0$ && $0.0$ && $W_{\mathrm{GAUSS}}=0.12$ \\
$\eta$ && SIN && $15$ && --- && $\eta\in[0.00,2.50]$
&& GAUSS && $1.5$ && $-15.0$ && $W_{\mathrm{GAUSS}}=0.12$ \\
$\xi$ && SIN && $15$ && --- && $\xi\in[0.00,2.50]$
&& GAUSS && $1.5$  && $-15.0$ && $W_{\mathrm{GAUSS}}=0.12$ \\
$Q_{0x}$ && HO && $15$ && $Q_{0x}^{(0)}$ && $500.0$ cm$^{-1}$, $63.546$ dalton 
&& HO && $Q_{0x}^{(0)}$ && $0.0$ && $500.0$ cm$^{-1}$, $63.546$ dalton \\
$Q_{0y}$ && HO && $15$ && $Q_{0y}^{(0)}$ && $500.0$ cm$^{-1}$, $63.546$ dalton
&& HO && $Q_{0y}^{(0)}$ && $0.0$ && $500.0$ cm$^{-1}$, $63.546$ dalton \\
$Q_{0z}$ && HO && $15$ && $Q_{0z}^{(0)}$ && $500.0$ cm$^{-1}$, $63.546$ dalton 
&& HO && $Q_{0z}^{(0)}$ && $0.0$ && $500.0$ cm$^{-1}$, $63.546$ dalton \\
$\{Q_{i\mu}\}_{i=1}^{8}$, $\mu=x,y,z$ && HO && $10$ &&
$\{Q_{i\mu}^{(0)}\}_{i=1}^{8}$ && $500.0$ cm$^{-1}$, $63.546$ dalton
&& HO && $\{Q_{i\mu}^{(0)}\}_{i=1}^{8}$ && $0.0$ && $500.0$ cm$^{-1}$, $63.546$ dalton \\
$\{Q_{i\mu}\}_{i=9}^{20}$, $\mu=x,y,z$ && HO && $7$ &&
$\{Q_{i\mu}^{(0)}\}_{i=9}^{20}$ && $500.0$ cm$^{-1}$, $63.546$ dalton
&& HO && $\{Q_{i\mu}^{(0)}\}_{i=9}^{20}$ && $0.0$ && $500.0$ cm$^{-1}$, $63.546$ dalton \\
$\{Q_{i\mu}\}_{i=21}^{29}$, $\mu=x,y$ && HO && $5$ &&
$\{Q_{i\mu}^{0}\}_{i=21}^{29}$ && $500.0$ cm$^{-1}$, $63.546$ dalton
&& HO && $\{Q_{i\mu}^{0}\}_{i=21}^{29}$ && $0.0$ && $500.0$ cm$^{-1}$, $63.546$ dalton \\
$\{Q_{iz}\}_{i=21}^{29}$ && HO && $7$ &&
$\{Q_{iz}^{(0)}\}_{i=21}^{29}$ && $500.0$ cm$^{-1}$, $63.546$ dalton
&& HO && $\{Q_{iz}^{(0)}\}_{i=21}^{29}$  && $0.0$ && $500.0$ cm$^{-1}$, $63.546$ dalton \\
\hline
 \end{tabular}
   \label{tab:98d-num-prop}
    \end{sidewaystable}
    
\clearpage
\begin{sidewaystable}
 \caption{%
Comparisons of the concepts in the Berry phase effects on the nuclear
and electronic properties, together with these concepts associated with
electromagnetic (EM) field. The second column provides geometric concepts
that will be compared to various fields. These concepts are the Berry
connection, Berry curvature, and Berry phase, which arise in the parameter
space. The third and fourth columns list physical insights of these
concepts for the nuclear and electronic properties. The fifth column
lists the electromagnetic quantities or properties associated with
the geometric features. The rightmost column gives remarks of these
physical quantities or geometric properties.
}%
 \begin{tabular}{lcllllllllr}
  \hline
No. &~~& Feature \footnote{For clarity, the term ``Berry'' is omitted for the Berry
 connection, Berry curvature, and Berry phase.}  
 &~~& Nuclear &~~& Electron &~~& EM field &~~& Remarks \\
\hline
1 && connection &~~& $A_n(\mathfi{z})$ in Equation \eqref{eq:solution-of-se-001} 
           &~~& diagonal element of 
           &~~& analogous to the vector potential
\footnote{It is a concept in differential geometry but has similar
expression to the corresponding concepts in the classical EM field
leading to analogous phenomena.\label{foot:geom-em-field}}
           &~~& a gauge-dependent object   \\
&& && && electronic NACM 
\footnote{If the $j$-th electronic state $\upsilon_j(\mathbi{x},\mathbi{y})$
depends on electronic cooridnates $\mathbi{x}$ and parameters $\mathbi{y}$,
the elements of the electronic NACM satisfy
$\tau_{ij}^{\mathrm{elec}}(\mathbi{y})=\langle\upsilon_i\vert\nabla_{\mathbi{y}}\vert\upsilon_j
\rangle_{\mathbi{x}}$.\label{foot:elec-connection}} 
 &&  &~~& encoding system's dependence  \\
&& && && &&  &~~& on the parameters space  \\
2&&curvature  &~~& $F_n(\mathfi{z})$ in Equation \eqref{eq:chern-number}
           &~~& curl of $\tau_{nn}^{\mathrm{elec}}(\mathbi{y})$
\footnote{The Berry curvature of the $n$-th electronic state is
$\Omega_n(\mathbi{y})=\nabla_{\mathbi{y}}\times\tau_{nn}^{\mathrm{elec}}(\mathbi{y})$.
See also Footnote \textsuperscript{\ref{foot:elec-connection}}.}
           &~~& effective magnetic field in 
           &~~& a local geometric property of \\
&& && && && parameter space \textsuperscript{\ref{foot:geom-em-field}}
           &~~& parameter space encoding the  \\
&& && && &&  &~~& states' anholonomy under  \\
&& && && &&  &~~& adiabatic parameter variations  \\ 
3&&phase      &~~& nuclear wave function &~~& electronic wave function 
           &~~& geometric propertities in both &~~& geometric properties of \\ 
&& && && &&  classical polarization dynamics &~~& wave eigenstates   \\
&& && && &&  and quantum photonic system    &~~&  \\
4&&Chern number&& number of critical points &~~& number of CI points &~~& global properties of EM modes
\footnote{In general, the EM modes are defined for photons, or more pricisely bosons,
such as photon bands in periodic structures.}
           &~~& a topological invariant classifing \\
&& && && &&  &~~& the global topology of the system’s \\
&& && && &&  &~~& eigenstate bundle  \\ 
\hline
\end{tabular}
\label{tab:geom-comp}
\end{sidewaystable}

\clearpage
\section*{Figure Captions}

\figcaption{fig:cho}{%
A cartoon of the 3D CHO models, together with eigenvalues degeneracy
seams of the 3D CHO model at different mass parameters, where $m_1$,
$m_2$, and $m_3$ denote the masses of the three atoms while $\omega_1$,
$\omega_2$, and $\omega_3$ denote associated frequencies.
}%

\figcaption{fig:rex-pes}{%
2D surface plots and contour plots of $V_{\mathrm{3D-REX}}$ at (a)
$\theta=0.0$, (b) $\theta=\pi/4$, and (c) $\theta=\pi/2$, where the
parameters are given in Table \ref{tab:model} and the masses are all
$1.0$ dalton. The two axes of the horizontal plane are $r_{\mathrm{d}}$
and $r_{\mathrm{v}}$, and the vertical axis represents potential energy
(in eV). The surface and contour plots are displayed in the upper and
lower panels, respectively, in each subfigure. All of plots are computed
at $\alpha=\beta=1$. 
}%

\figcaption{fig:tot-phase}{%
Time-dependent phases (in radian) of the nuclear wave functions computed
by the present MCTDH calculations through (a) the 3D CHO model and (b)
the 3D REX model, where initial conditions and parameters are set to
similate different molecular systems. The black, red, blue, and green
lines represent the phase angle of CO/Cu, HCH$_3$/Au, H$_2$O, and H +
H$_2$, respectively. The horizontal axis represents propagation time
(in fs), the vertical axis indicates the phase angle.
}%

\figcaption{fig:phase}{%
Time-dependent phases (in radian) of the nuclear wave functions computed
by the present MCTDH calculations through the 3D CHO model, where initial
conditions and parameters are set to similate different molecular systems,
including (a) CO/Cu, (b) HCH$_3$/Au, (c) H$_2$O, and (d) H + H$_2$. The
horizontal axis represents propagation time (in fs), the left vertical
axis indicates the phase angle (in radian), and the right vertical axis
(in red words) shows the time-dependent expectation of $\eta$. The
parameters of the 3D CHO model are given in Tables \ref{tab:mass-model}
and \ref{tab:model}. For clarity, in each subfigure we show only the
time period in which the phase changes. In each subfigure, the black
circles represent the time-dependent phase values (referred to the
left axis). The red line represent the time-dependent expectation
$\langle\eta\rangle(t)=\langle\Psi(t)\vert\eta\vert\Psi(t)\rangle/\langle\Psi(t)\vert\Psi(t)\rangle$
(referred to the right axis). The blue symbol $\times$ means the
critical point $\eta_c=\sqrt{1+m_1/\max(m_2,m_3)}$ (referred to the
right axis), where the phase should change.
}%

\begin{figure}[h!]
 \centering
  \subfigure[\quad Time-dependent phase of the 3D CHO model for H + H$_2$]{
   \includegraphics[width=18cm]{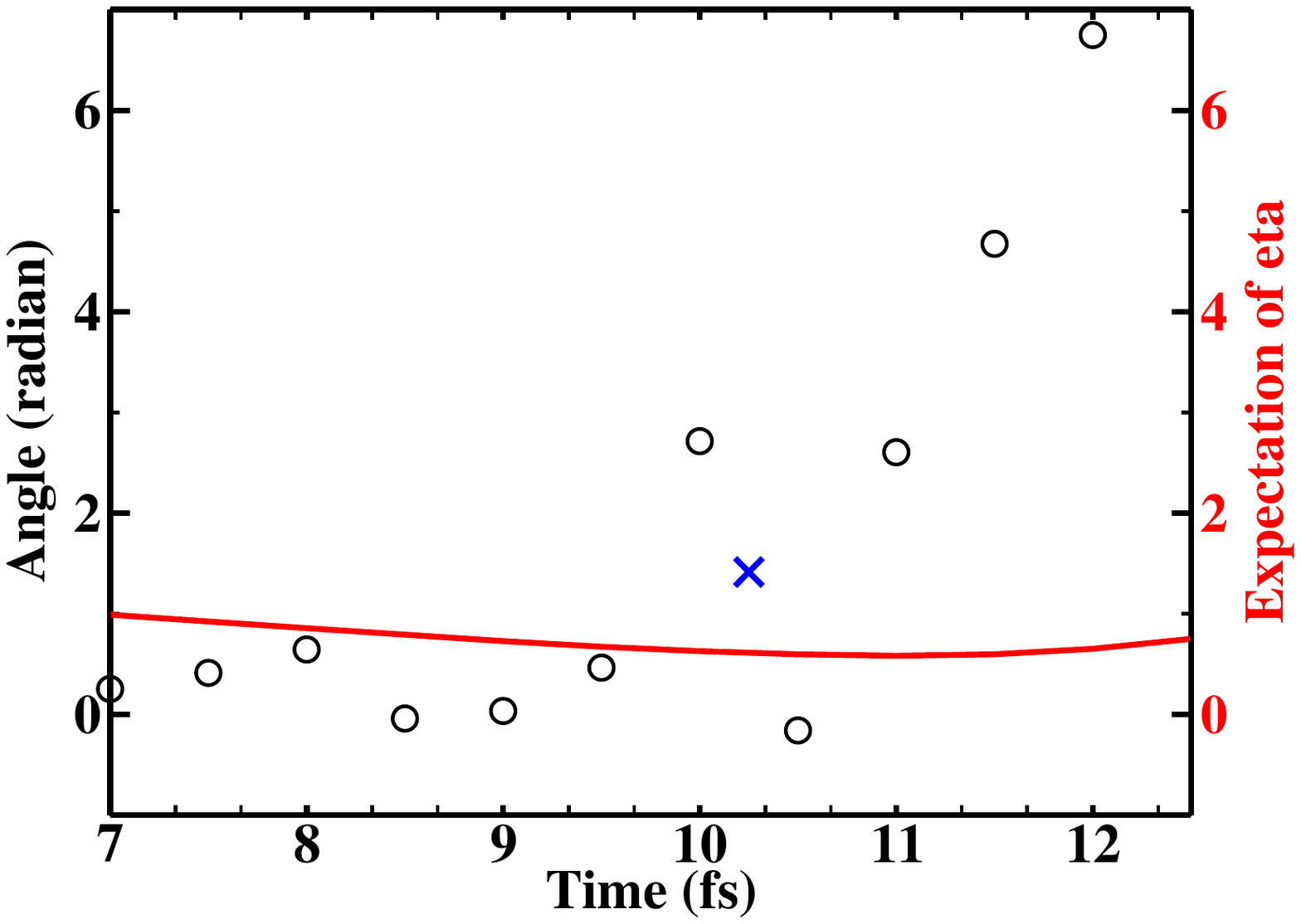}
}  
     \caption{\figfoot}
      \label{fig:phase}
       \end{figure}

\figcaption{fig:phase-prop}{%
Same as Figure \ref{fig:phase} except for the 3D REX model, which is
adopted to model scattering processes of (a) CO/Cu, (b) HCH$_3$/Au, 
c) H$_2$O, and (d) H + H$_2$. Moreover, since the time-dependent phase
of H + H$_2$ changes dramatically, one cannot clearly find transition
of the phase angle and thus the time where the critical point $\sqrt{2}$
occurs. For clarity, the critical point is shown by a blue dashed line
in subfigure (d).
}%

\figcaption{fig:geom-ben}{%
Geometry of the Cu(100) surface in the numerical demonstration on the
present theory in surface scattering. At each the first and third layers of
the model surface, a total of nine surface atoms are uniformly
distributed across a $3\times3$ grid layout and represent by black
circles, while the atoms of the seocnd layer are given by symbols
$\times$. The center-of-mass (COM) of the molecule is represented by
red circle. The surface plane is the $x$-$y$ plane and hence the
original point is set at thetop atom, denoted by \#0. Moreover, the
surrounding atoms at the first layer are sequentially numbered \#1
through \#8 in a clockwise arrangement centered on atom \#0. The $x$
and $y$ axes are defined by the \#0-\#4 and \#0-\#2 connections,
respectively. The molecular COM is described by $\{R,\vartheta,z\}$,
where $R$ denotes the radial distance from the origin to the projected
point of the molecule and $\vartheta$ represents the polar angle of this
projected point. The chemical bonding between the surface atoms is
governed by the Morse potential. The distance between adjacent surface
atoms is denoted by $d$. By the same Morse potential, all of the surface
atoms are fixed at solid surface along negative $z$- and lateral-directions.
}%

\begin{figure}[h!]
   \centering
    \includegraphics[width=18cm]{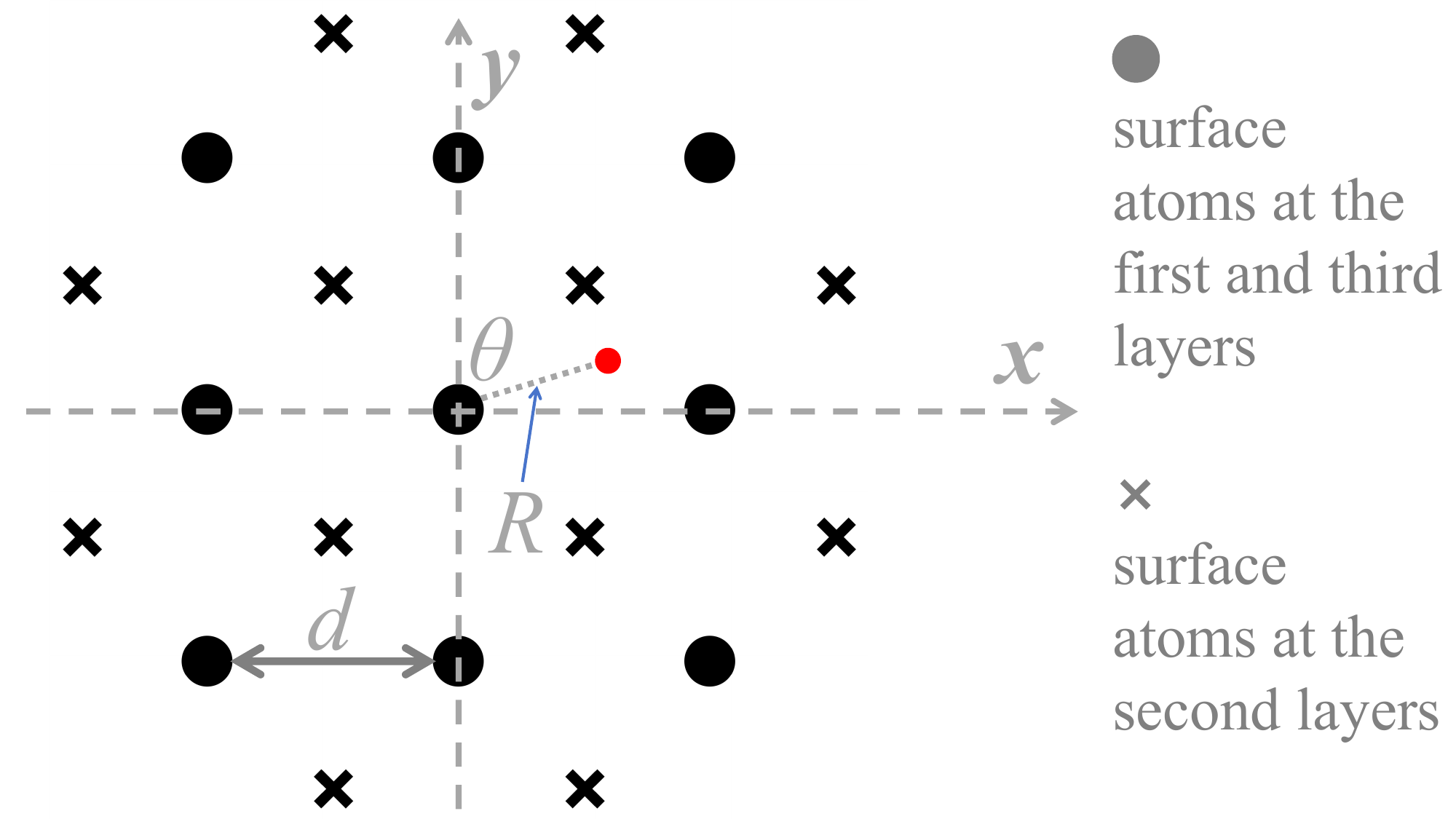}
     \caption{\figfoot}
      \label{fig:geom-ben}
       \end{figure}

\figcaption{fig:ml-tree-98d}{%
The ML-MCTDH wave function structure (called ML-tree structure) for the
present 98D calculations based on the Hamiltonian model given in Table
\ref{tab:model}. The maximum depth of the ML-tree is six layers, and
the first layer separates the moelcular coordinates from those of the
surface atoms, as shown in Figure \ref{fig:geom-ben}. The number of
SPFs are also given. The numbers of primitive basis sets to represent
SPFs of the deepest layer are given next to the lines connecting with
the squares (see also Table \ref{tab:num-prop}). For simple, subfigure
(a) only shows the first $35$ coordinates, while subfigure (b) shows
the other $63$ coordinates that should locate at the ML-tree structure
in the red box of subfigure (a).
}%

\begin{figure}[h!]
   \centering
    \subfigure[\quad ML-tree for the other part]{%
     \includegraphics[width=23cm,angle=90]{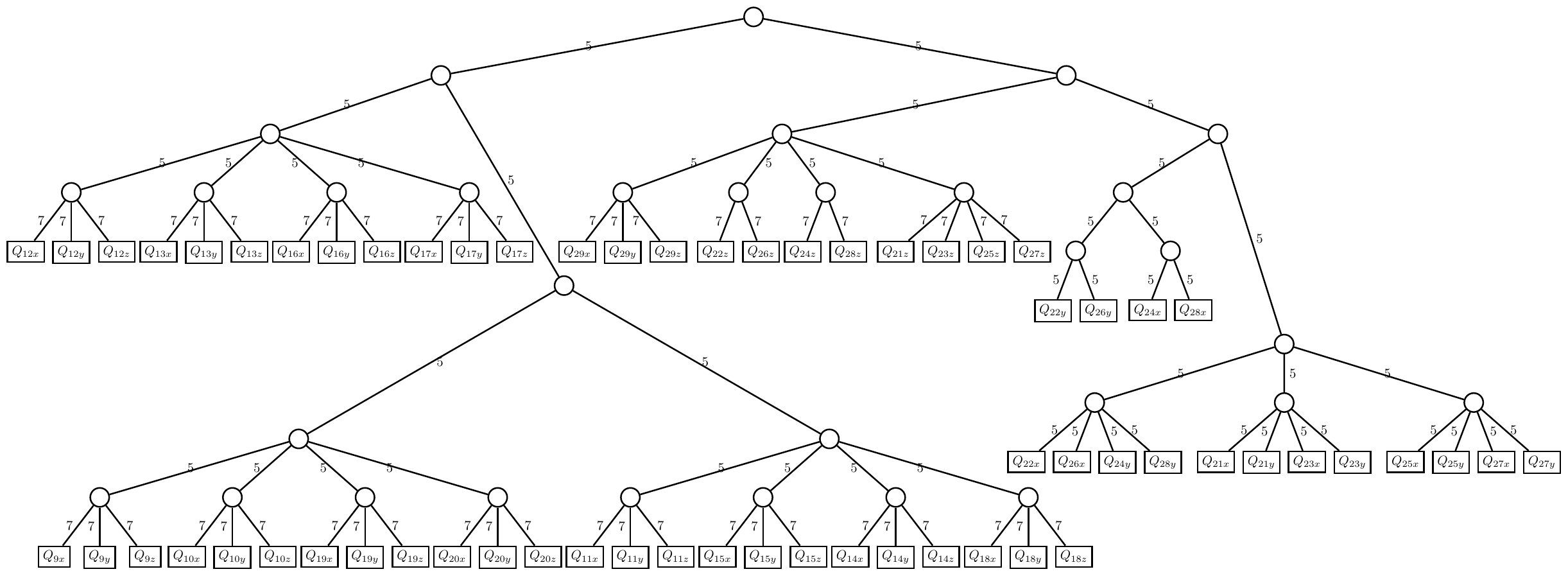}}
     \caption{\figfoot}
      \label{fig:ml-tree-98d}
       \end{figure}

\figcaption{fig:98d-co-cu-dyn}{%
Same as Figure \ref{fig:phase} except for the 98D CHO model, which is
adopted to model scattering processes of CO/Cu. The atomic arrangement
and the ML-tree of the 98D wave function are given in Figures \ref{fig:geom-ben}
and \ref{fig:ml-tree-98d}, respectively.
}%

\figcaption{fig:expert}{%
Ideal experiment to verify the sign change of the molecular wave
function as indicated by Equation \eqref{eq:adi-dia-rep-014}, drawn
inspiration from experiments reported by Yang and co-workers
\cite{xie20:767} and Beck and co-workers \cite{rei25:962}. It can be
carried out using a molecule-beam apparatus based on rather light
molecule, say hydrogen or methane, which scatters from a metal surface,
say gold, held at given surface temperature. A pump laser should be
adopted for selecting rovibrational states of the moelcule.
The surface should be rotatable around the center of the scattering
target region, so that possible to vary the incident angle. The
scattered molecule was detected by a tagging laser combined with a
bolometer detector to permit state–resolved population measurements
of the scattered molecule. The incident angle and scattering angle,
defined with respect to the surface normal, are independently variable.
}%

 \clearpage
       
\clearpage
  \begin{figure}[h!]
   \centering
    \subfigure[\quad Surface plot and contour plots of the PES for the 3D REX model at $\theta=0.0$]{
     \includegraphics[width=18cm]{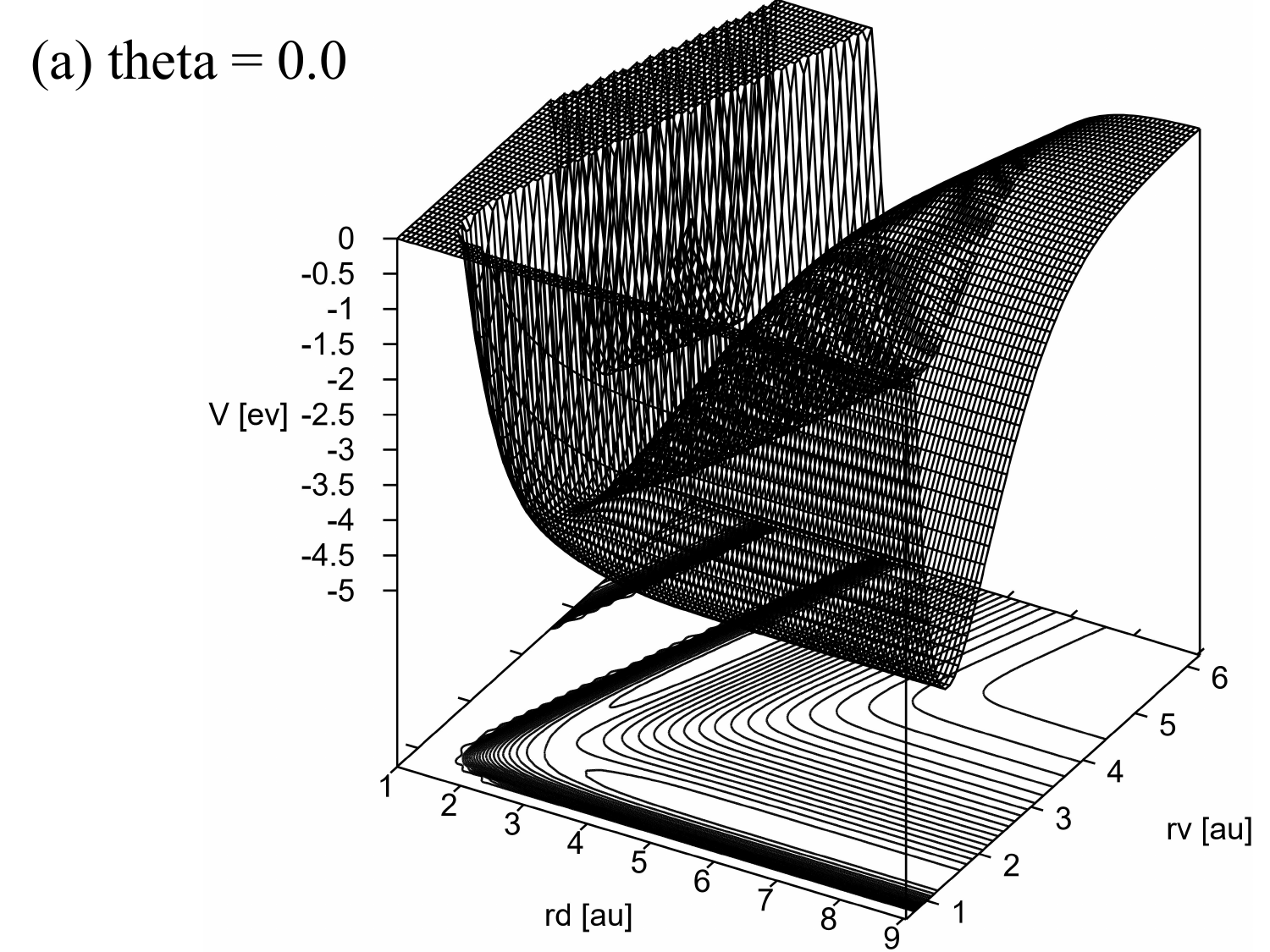}}
      \end{figure}
\clearpage
\begin{figure}[h!]
   \centering
    \subfigure[\quad Surface plot and contour plots of the PES for the 3D REX model at $\theta=\pi/4$]{
     \includegraphics[width=18cm]{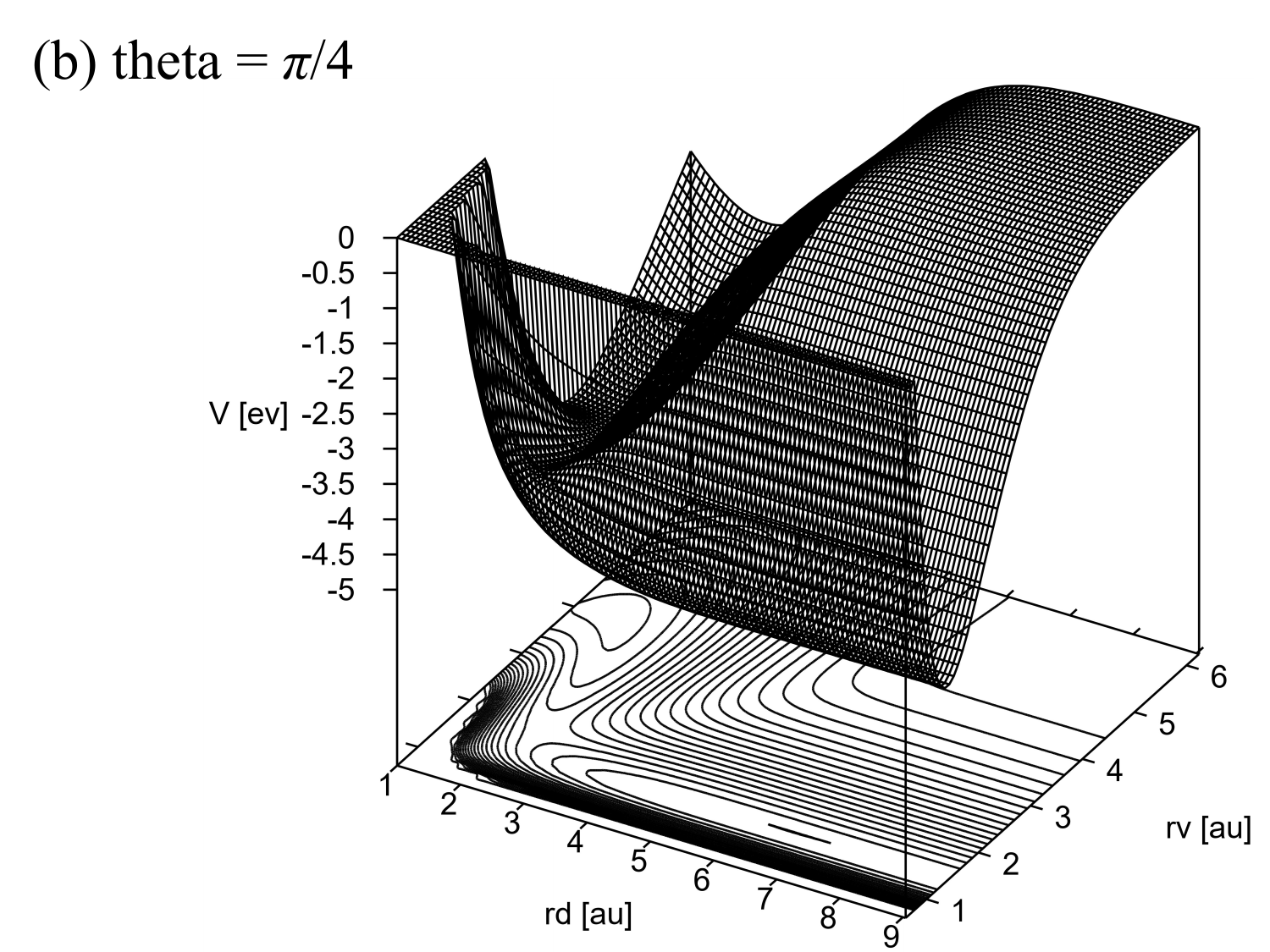}}
      \end{figure}
   
\clearpage
\begin{figure}[h!]
 \centering       
  \subfigure[\quad Phase angle of the 3D CHO model]{
   \includegraphics[width=18cm]{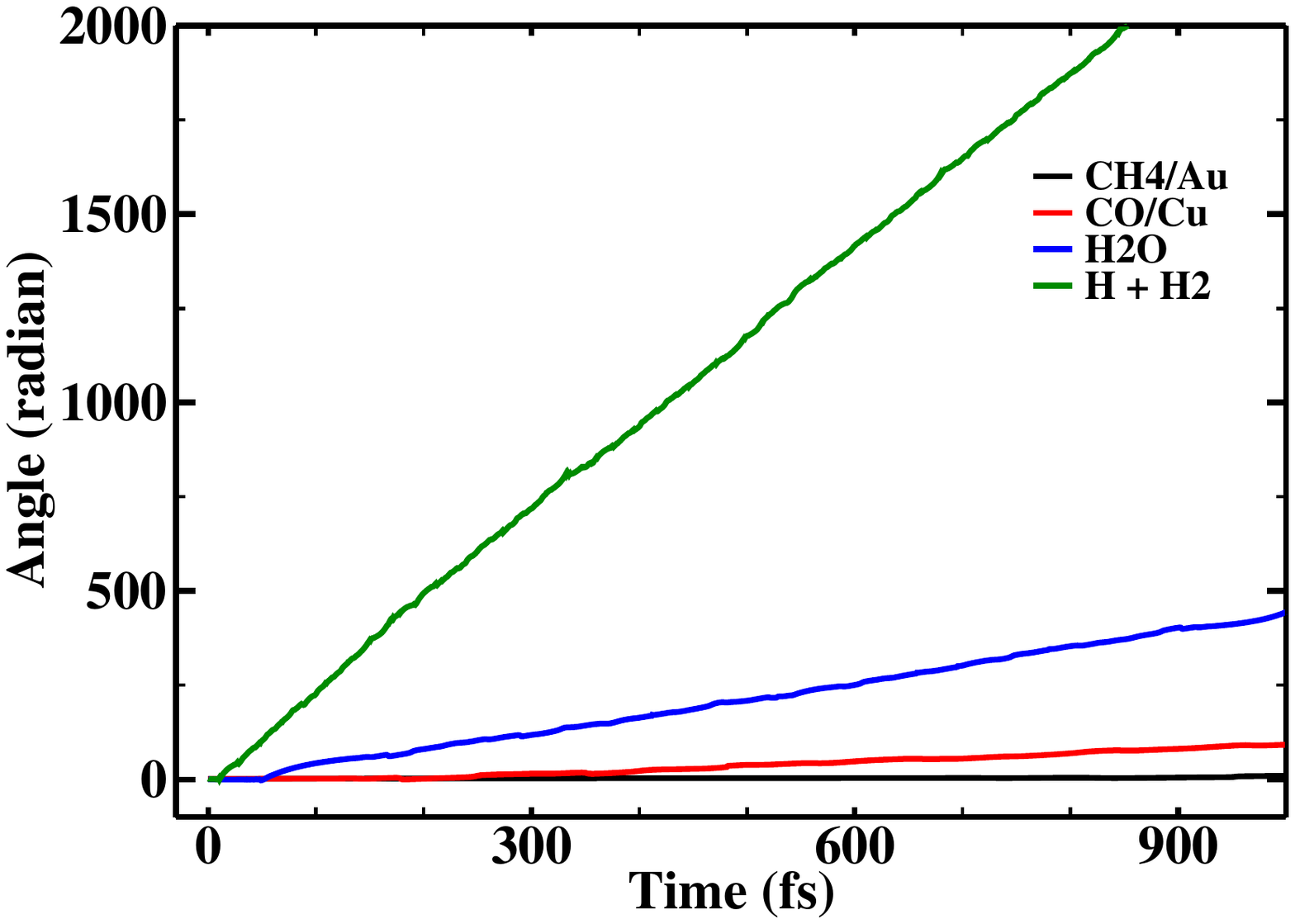}
}
       \end{figure}
      
 \clearpage
  \begin{figure}[h!]
   \centering
    \subfigure[\quad Time-dependent phase of the 3D CHO model for CO/Cu]{
     \includegraphics[width=18cm]{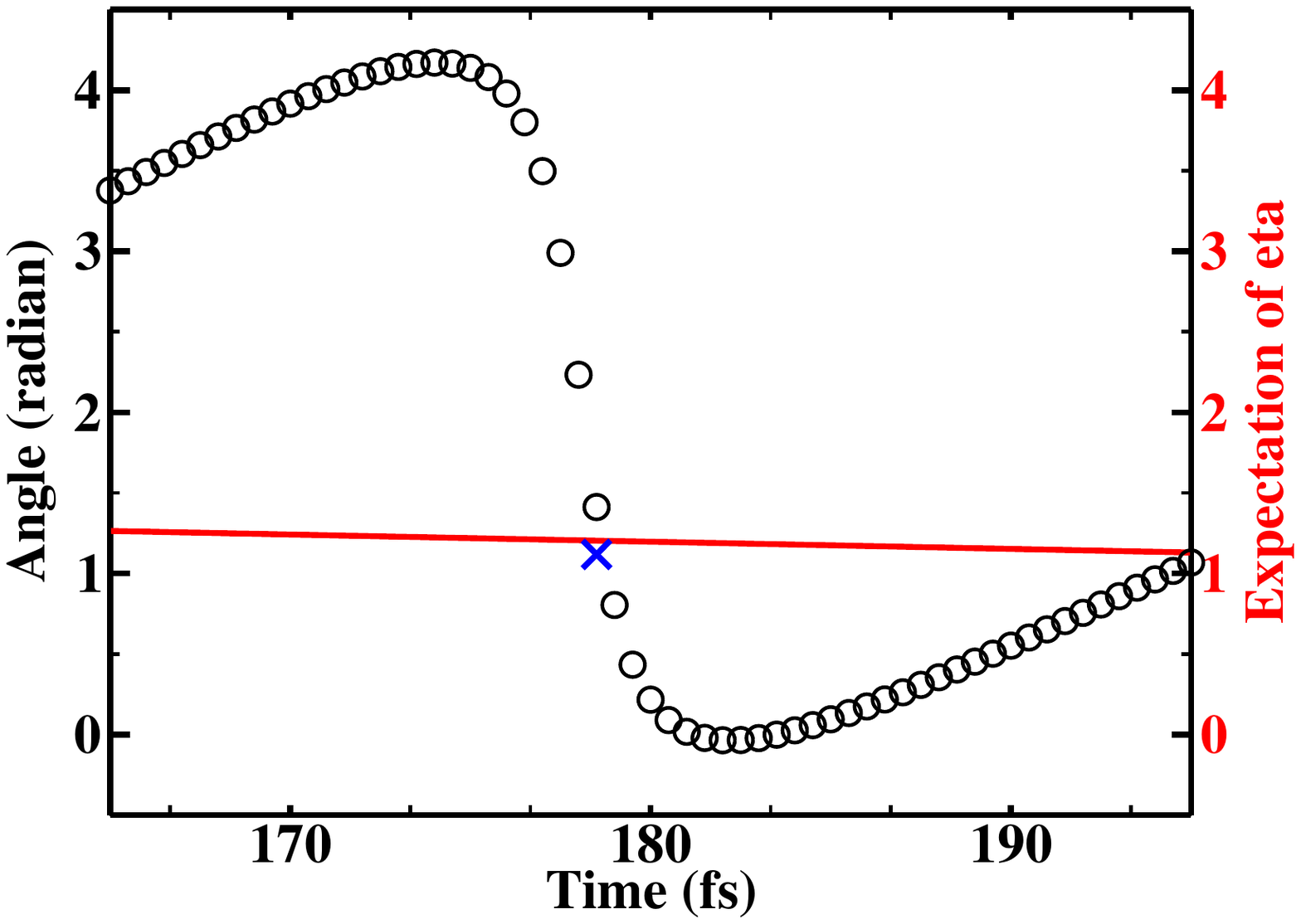}
}
       \end{figure}
  \begin{figure}[h!]
   \centering
    \subfigure[\quad Time-dependent phase of the 3D CHO model for CH$_4$/Au]{
     \includegraphics[width=18cm]{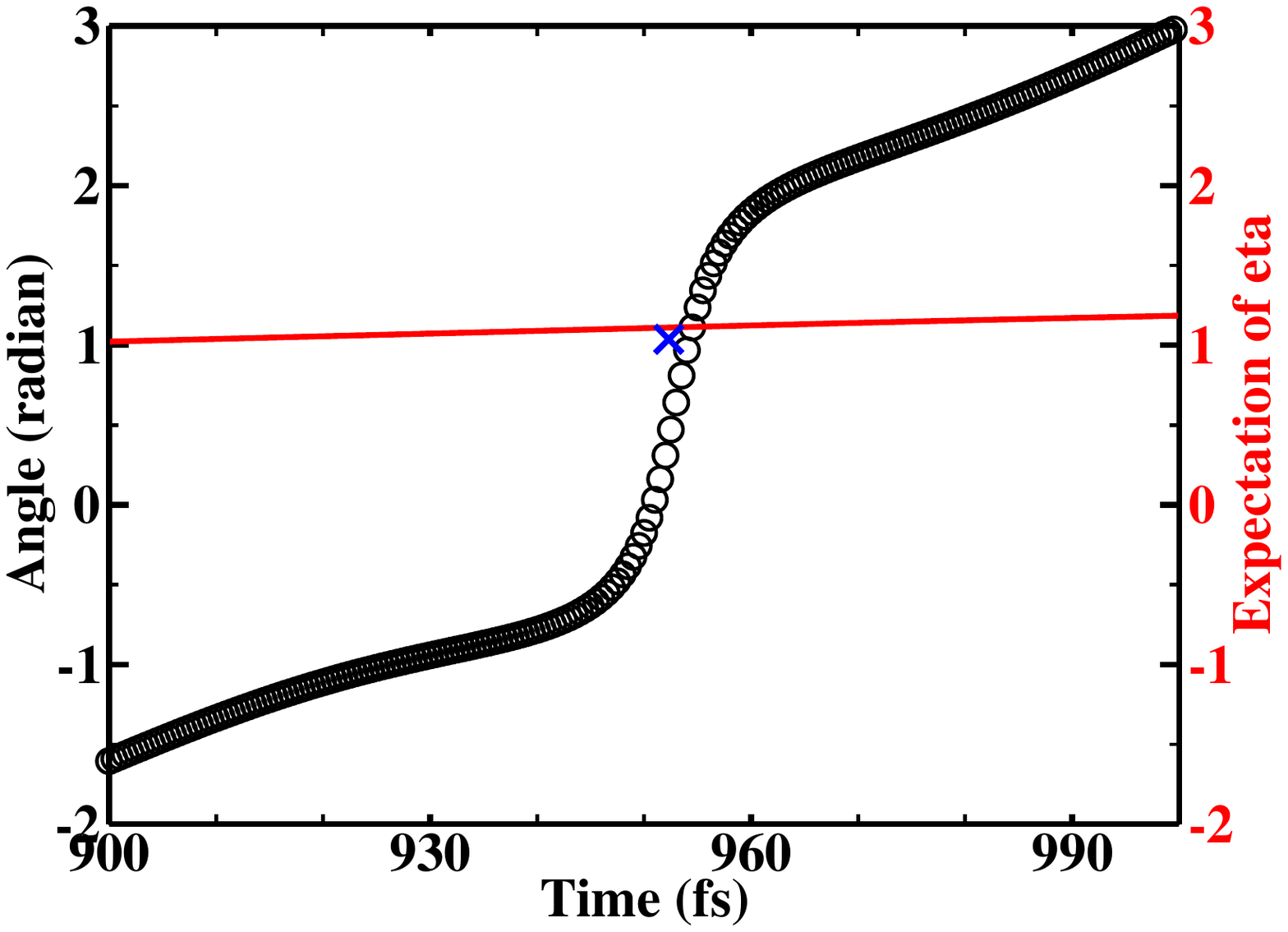}
}
       \end{figure}
\begin{figure}[h!]
 \centering       
  \subfigure[\quad Time-dependent phase of the 3D CHO model for H$_2$O]{
   \includegraphics[width=18cm]{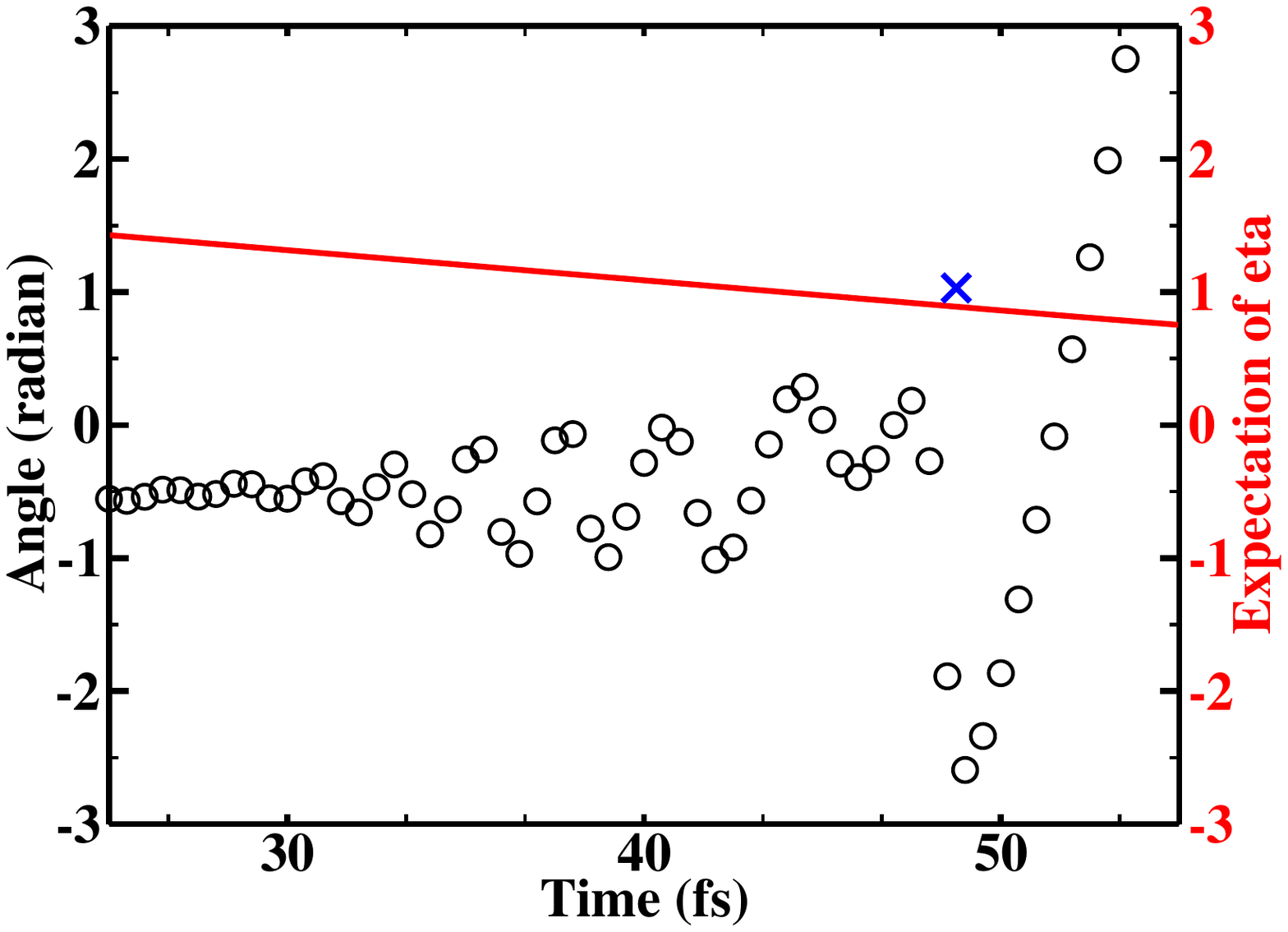}
}
       \end{figure}
       
 \clearpage
  \begin{figure}[h!]
   \centering
    \subfigure[\quad Time-dependent phase of the 3D REX model for CO/Cu]{
     \includegraphics[width=18cm]{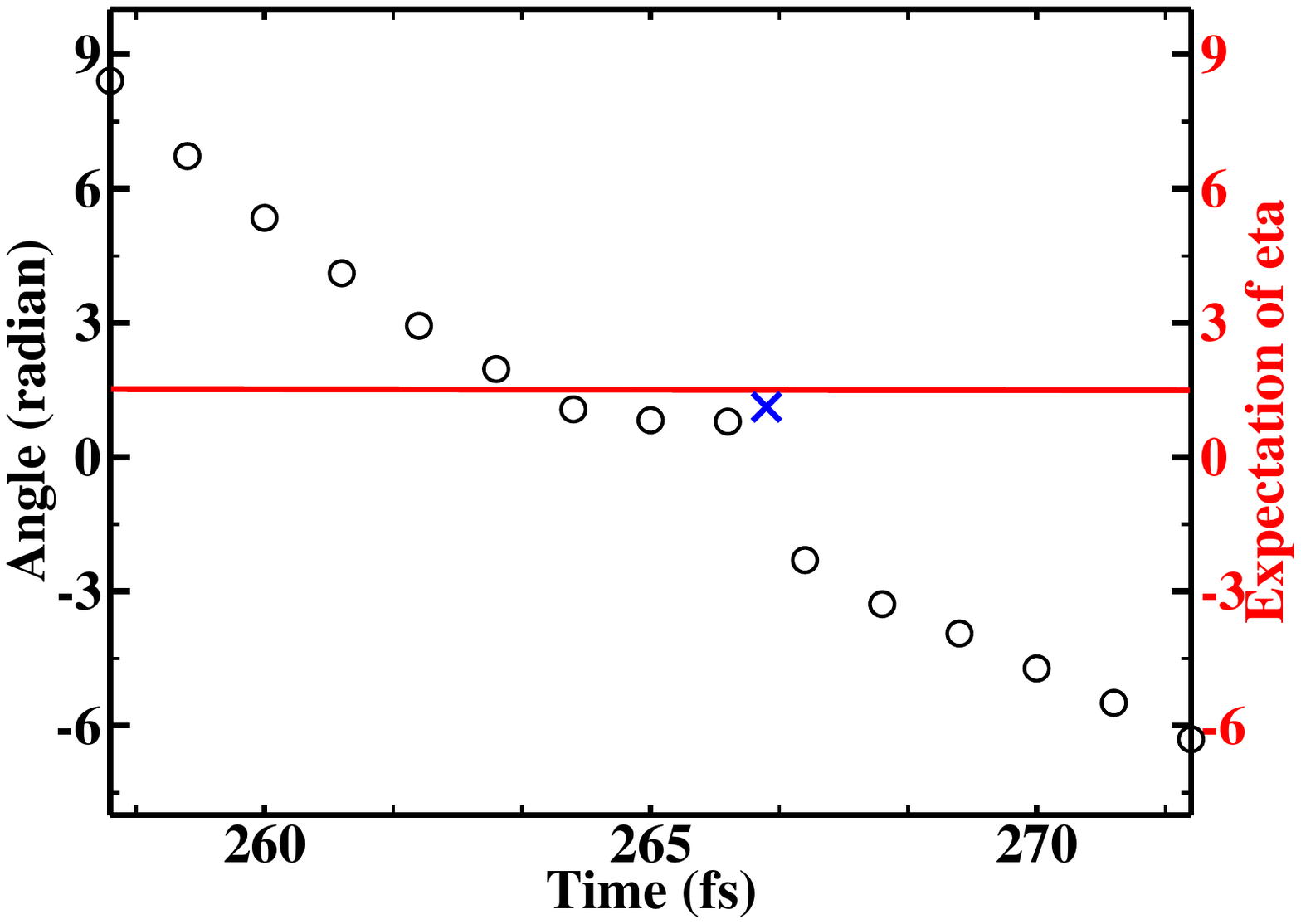}
}
       \end{figure}
  \begin{figure}[h!]
   \centering
    \subfigure[\quad Time-dependent phase of the 3D REX model for CH$_4$/Au]{
     \includegraphics[width=18cm]{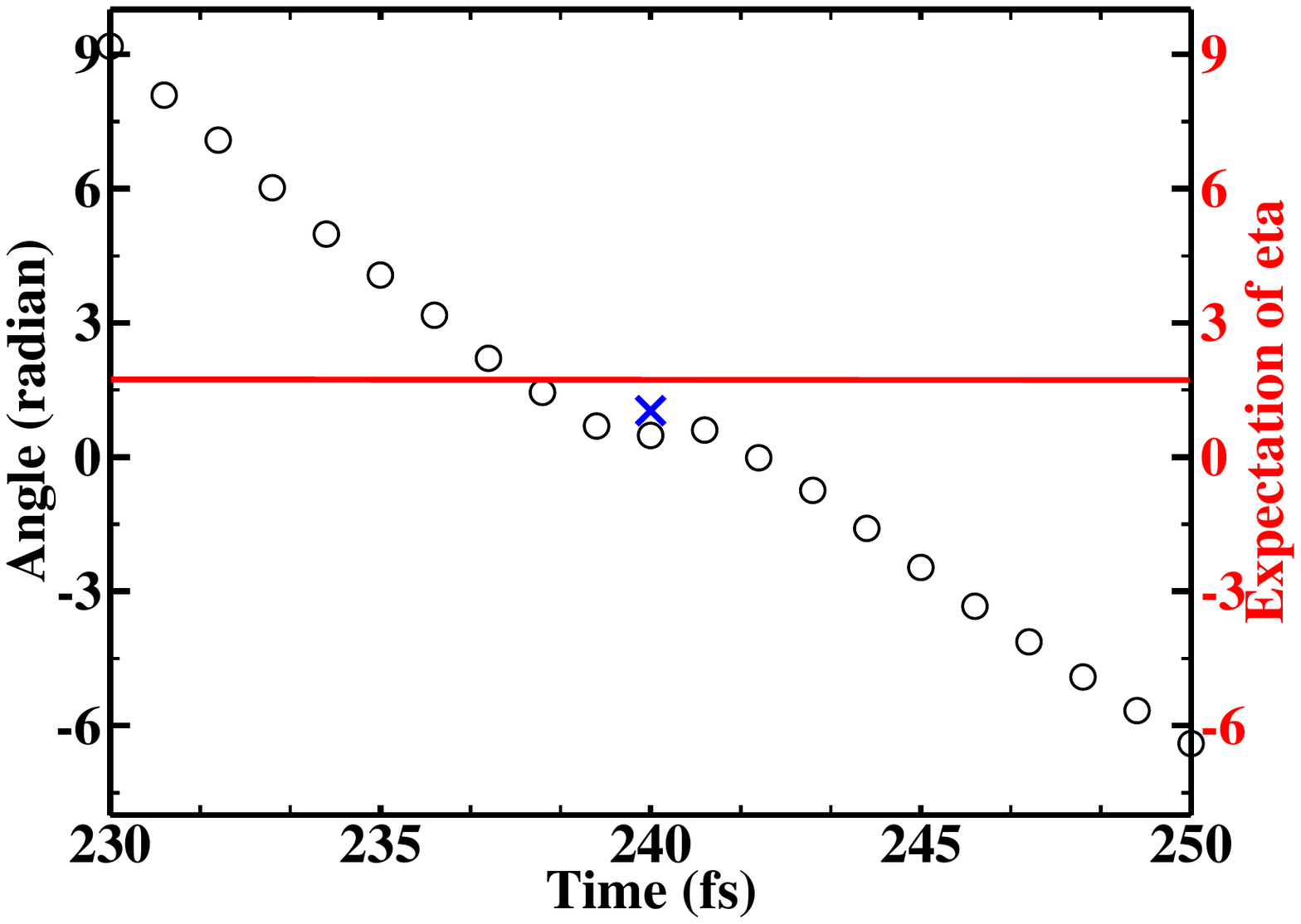}
}
       \end{figure}
\begin{figure}[h!]
 \centering       
  \subfigure[\quad Time-dependent phase of the 3D REX model for H$_2$O]{
   \includegraphics[width=18cm]{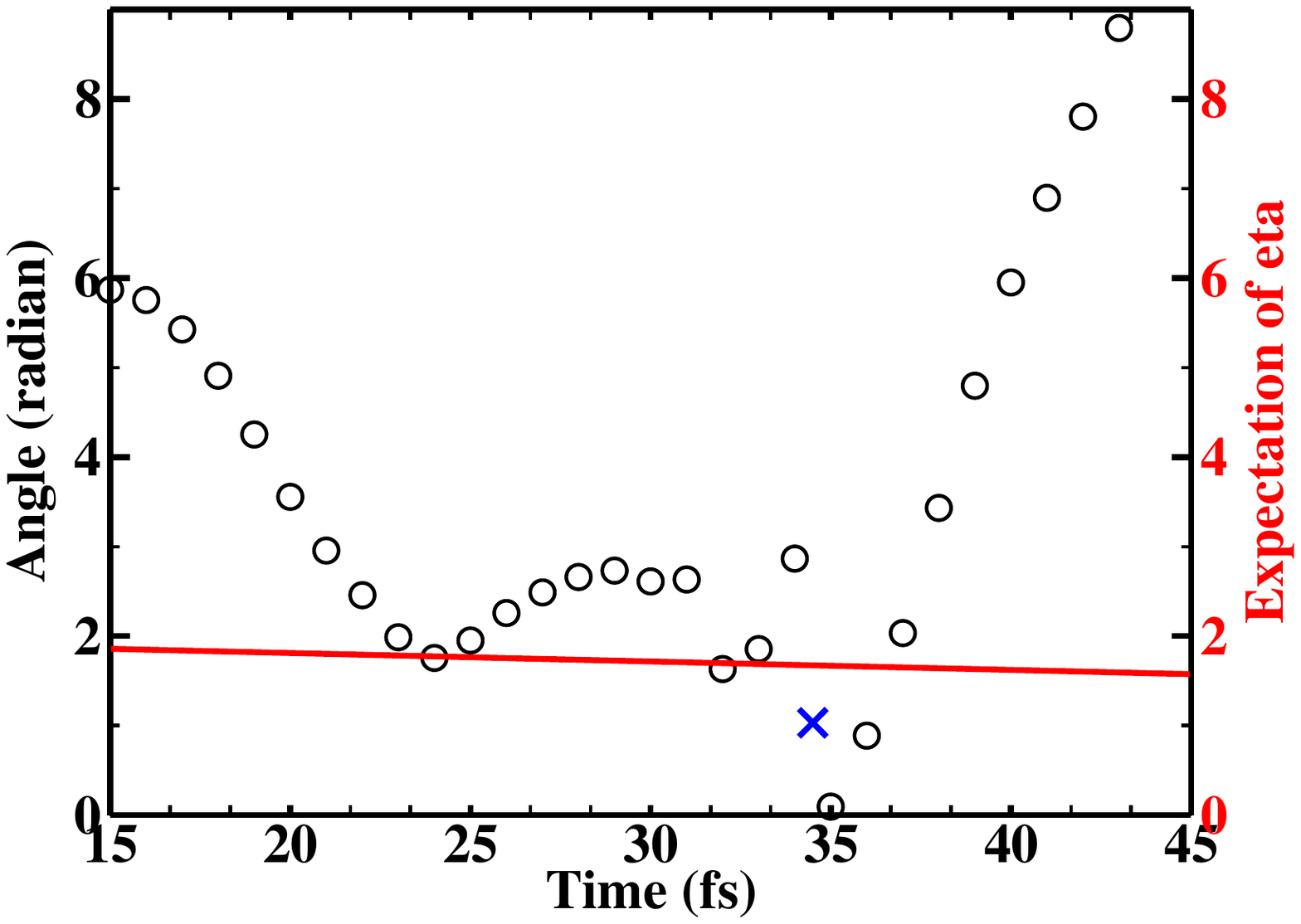}
}
       \end{figure}

\clearpage
       
\clearpage
  \begin{figure}[h!]
   \centering
    \subfigure[\quad ML-tree of the entire wave function]{%
     \includegraphics[width=23cm,angle=90]{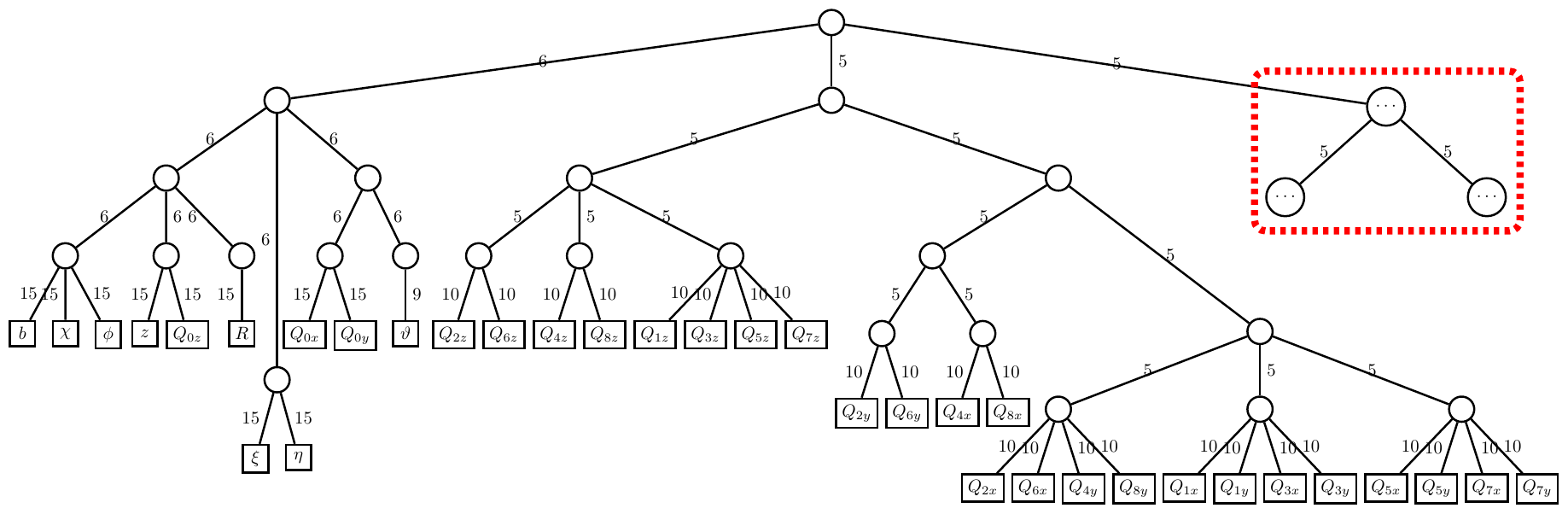}}
      \end{figure}
 \clearpage
       
\clearpage

 \clearpage
       

\end{document}